\RequirePackage{fix-cm}
\documentclass[smallextended]{svjour3}
\usepackage{graphicx, amsmath, amssymb, enumitem}
\usepackage[spaces]{grffile}

\newcommand{\pmat}[2]{\def\arraystretch{#2}\begin{pmatrix}#1\end{pmatrix}}%\mat{input}{stretch amount}
\newcommand{\bmat}[2]{\def\arraystretch{#2}\begin{bmatrix}#1\end{bmatrix}}%\mat{input}{stretch amount}
\newcommand{\re}{\,\textrm{Re}}
\newcommand{\im}{\,\textrm{Im}}
\newcommand{\sgn}{\,\textrm{sgn}}

\begin{document}
	\title{Dynamical response of a system of passive or active swimmers to time-periodic forcing}
	
	\author{Michael Wang \and Alexander Y. Grosberg}
	
	\institute{M. Wang \\ \email{mw3189@nyu.edu} \\ A. Y. Grosberg \\ \email{ayg1@nyu.edu \\ Department of Physics and Center for Soft Matter Research, New York University, 726 Broadway, New York, New York 10003, USA}}
	
	\maketitle
	
	\begin{abstract}
		The presence of active forces in various biological and artificial systems may change how those systems behaves under forcing.  We present a minimal model of a suspension of passive or active swimmers driven on the boundaries by time-dependent forcing.  In particular, we consider a time-periodic drive from which we determine the linear response functions of the suspension.  The meaning of these response functions are interpreted in terms of the storage and dissipation of energy through the particles within the system.  We find that while a slowly driven active system responds in a way similar to a passive system with a re-defined diffusion constant, a rapidly driven active system exhibits a novel behavior related to a change in the motoring activity of the particles due to the external drive.
		
	\end{abstract}
	
	\section{\label{sec:Introduction}Introduction}
		Active matter has been of great interest due to its use in understanding a wide range of biological and artificial out-of-equilibrium systems \cite{Marchetti et al,Bechinger et al}.  Many active systems studied consist of individual particles that consume energy locally and generate independent motion.  Examples of these self-propelled particles include bacteria (e.g.\ \textit{E.coli} \cite{Berg et Brown}); their artificial counterparts, micron-sized catalytic swimmers \cite{Palacci et al,Paxton et al}; and molecular motors \cite{Astumian,Julicher et al}.
		
		Two quantities that characterize the motion of self-propelled particles is a propulsion force or velocity and most importantly, a persistence time for the direction of propulsion.  On short time scales, their motion is ballistic-like, while on long time scales in free space, they undergo random walks and effectively diffuse, much like passive Brownian particles undergoing thermal diffusion.  However, despite the similarities with passive particles on long time scales, it is known that a nonzero persistence time combined with interactions with an environment, for example obstacles and other particles, can lead to emergent out-of-equilibrium behaviors: liquid-gas phase separation in the absence of attractive interactions \cite{Cates and Tailleur,Redner et al,Fily and Marchetti,Wysocki et al,Bialke et al}, preferential motion of bacteria in one direction through funnel-shaped gates \cite{Galajda et al}, and bacteria-powered microscopic ratchets, gears, or motors \cite{Di Leonardo et al,Vizsnyiczai et al}.  In addition, the persistent nature of active particles---and more generally, systems containing components driven by non-thermal noise---has noticeable effect on the dynamic response of those systems to mechanical and chemical perturbations \cite{Sokolov et al,Rafai et al,Hatwalne et al,Turlier et al,Chu et al,Fodor et al,Bi et al,Caprini et al,Solon et al,Marconi et al,Sheshka et al}.  What has been less studied is how the details of the diffusive and ballistic movements of passive and active particles within a larger system may affect that system's response to external forcing.
		
		In this paper, we present a solvable minimal model of a suspension of non-interacting passive or active particles driven periodically at the boundaries.  We extract the response functions of the system, which relate the external forcing to the behavior of the suspension, and consider how the particles store and dissipate energy from the drive.  We find that the diffusivities of the particles play an important role in how the suspension responds to forcing on different time-scales.  We observe that on short time-scales, persistent particles indeed exhibit a response different from that of diffusive particles.
		
	\section{\label{sec:Model}Model}
		\begin{figure}
			\centering
			\includegraphics[scale=0.4]{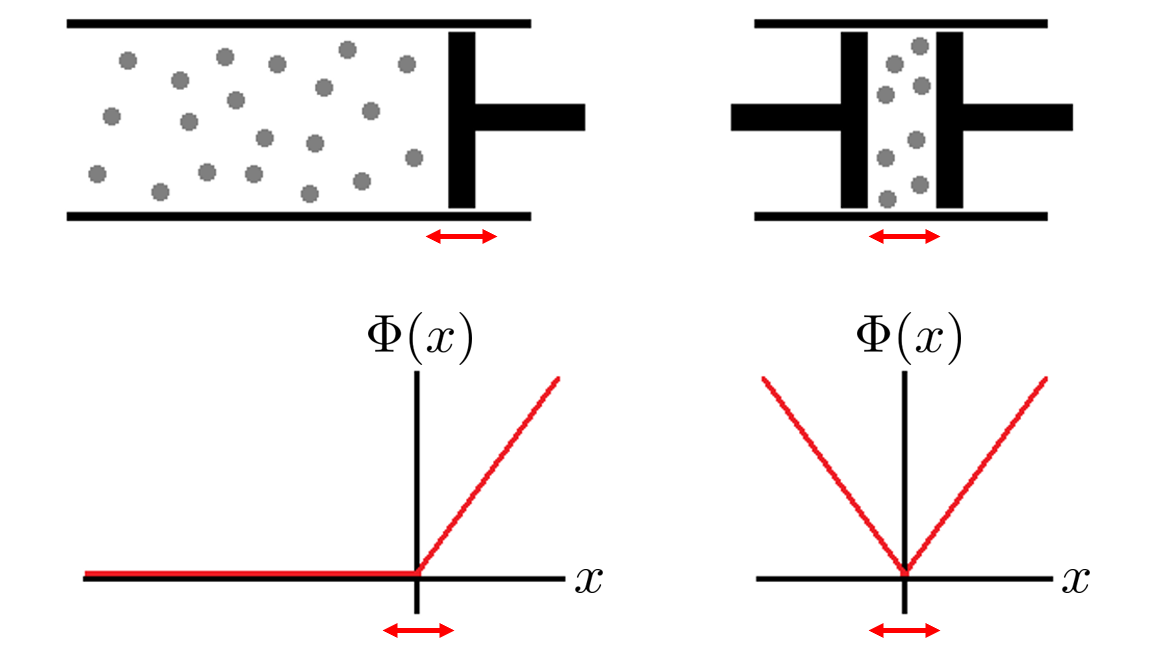}
			\caption{Model of the system.  \textbf{Left:} Large bulk of particles confined by a piston and the corresponding ramp potential (Eq.\ (\ref{ramp potential})).  \textbf{Right:} Particles trapped and transported between two co-moving pistons represented by a V-shaped potential (Eq.\ (\ref{v-shaped potential})).  The red double arrows indicate movement of the pistons and potentials due to drive.}
			\label{Model}
		\end{figure}
	
		To model an externally forced suspension of passive or active particles, we consider two scenarios (Fig.\ \ref{Model}): a uniform bulk of particles confined by a single piston and particles trapped between two pistons.  To make this model analytically tractable, we restrict ourselves to 1D and describe the pistons as linear potentials.  In the first case of a single piston, the confining potential is given by a ramp potential
		\begin{equation}
			\Phi_{\textrm{ramp}}(x)=
			\begin{cases}
				0, & x<0\\
				fx, & x\ge0
			\end{cases}.
			\label{ramp potential}
		\end{equation}
		Here, $f$ is the force experienced by the particles when they enter the ramp region.  In the second case of two pistons, the confining potential is given by a V-shaped potential
		\begin{equation}
			\Phi_{\textrm{V-shaped}}(x)=f\left|x\right|.
			\label{v-shaped potential}
		\end{equation}
		The presence of a bulk, or lack thereof, affects the response of these particles to an external time-dependent forcing.  A time-dependent forcing can be realized by moving the positions of the pistons according to some protocol $a(t)$, or mathematically, by replacing $\Phi(x)$ with $\Phi(x-a(t))$.  In this paper, we consider a time-periodic drive given by $a(t)=a\sin\omega t$.  Note that we assume the pistons are completely permeable to fluid such that the drive does not generate fluid flow and the only change to the local density of particles is through direct interaction with the potentials.
		
		The particles we study here are non-interacting passive particles with diffusivity $D$ and run-and-tumble (RnT) particles with propulsion velocity $\pm v$ and tumble rate $\alpha$.  The particles are assumed to be overdamped with mobility $\mu$ in the static fluid.  For passive particles, the density $\rho(x,t)$ evolves according to the usual advection-diffusion equation
		\begin{equation}
			\frac{\partial\rho}{\partial t}=-\frac{\partial}{\partial x}\Big[{-\mu\Phi'(x-a(t))\rho}\Big]+D\frac{\partial^2\rho}{\partial x^2}
			\label{eq:diffusion pde lab}
		\end{equation}
		while for RnT particles in 1D, the densities $\rho_{\pm}(x,t)$ of left ($+$) and right ($-$) moving particles evolve according to \cite{Schnitzer}
		\begin{align}
			\frac{\partial\rho_{\pm}}{\partial t}=-\frac{\partial}{\partial x}\Big[(\pm v-\mu\Phi'(x-a(t)))\rho_{\pm}\Big]-\alpha\rho_{\pm}+\alpha\rho_{\mp}.
			\label{eq:rnt pde lab}
		\end{align}
		The first term on the right-hand side is responsible for drift due to self-propulsion and external forces while the second and third terms capture the transitions or tumbles between the two propulsion directions.  We are interested in how perturbations to the local density influences the response due to external forcing.
		
	\section{\label{sec:Calculation}Calculation}
		\subsection{\label{subsec:Setup}Co-moving frame and dimensionless parameters}
			It is more convenient to solve the model described in Sec.\ \ref{sec:Model} in the rest frame of the pistons.  Transforming to said frame using $y=x-a\sin\omega t$ introduces a fictitious drift $-\dot{a}(t)=-a\omega\cos\omega t$ (see Appendix \ref{app:Solution Details} for the resulting PDEs).  
			
			We pick the characteristic time scale to be $\tau_{\textrm{drive}}=1/\omega$, the time scale of the drive.  Consequently, we pick the characteristic length scale to be the root mean-squared displacement of the particles in that time $\tau_{\textrm{drive}}$.  
			
			$\bullet$  For passive particles, that length scale is $l_{\textrm{diffusion}}=\sqrt{D\tau_{\textrm{drive}}}=\sqrt{D/\omega}$.  With these choices, Eq.\ (\ref{eq:diffusion pde lab}) becomes
			\begin{equation}
				\frac{\partial\rho}{\partial\tilde{t}}=-\frac{\partial}{\partial\tilde{y}}\left[\left(-\epsilon\cos\tilde{t}-\gamma\tilde{\Phi}'(\tilde{y})\right)\rho\right]+\frac{\partial^2\rho}{\partial\tilde{y}^2},
				\label{eq:diffusion pde rest dim}
			\end{equation}
			where the rescaled potential is $\tilde{\Phi}=\Phi/f$.  The dimensionless parameters are defined as
			\begin{equation}
				\epsilon=\sqrt{\frac{a^2\omega}{D}},\ \ \ \gamma=\sqrt{\frac{\mu^2f^2}{\omega D}}.
				\label{eq:def passive params}
			\end{equation}
			It is useful to note that $\epsilon$ and $\gamma$ can be written as the ratios $a/l_{\textrm{diffusion}}$ and $l_{\textrm{diffusion}}/l_{\textrm{penetration}}$, respectively.  Here, $l_{\textrm{penetration}}=D/\mu f$ (alternatively $k_BT/f$ for thermal particles) is approximately the maximum distance a diffusing particle penetrates into a linear potential region.
			
			We want to study the linear response of the system to external drive, that is, the response which is linear in the drive amplitude $a$.  We claim that the correct way to form the dimensionless criteria for determining the ``smallness'' of the amplitude $a$ is the above defined $\epsilon=a\sqrt{\omega/D}=a/l_{\textrm{diffusion}}$.  This is clear physically because at small drive frequencies $\omega$, even a relatively large drive amplitude $a$ corresponds to a gentle drive, as the system remains very close to steady-state at all times.  We therefore will perform expansions linear in $\epsilon\ll1$.  
			
			The parameter $\gamma$ characterizes how far the particles climb up or down the confining potentials and hence how much of the potentials they can explore over each cycle.  If $\gamma\gg1$ ($l_{\textrm{diffusion}}\gg l_{\textrm{penetration}}$), the particles have sufficient time to diffuse over the entirety of their confinement up to the penetration depth.  We call this the ``slow'' drive regime.  On the other hand, if $\gamma\ll1$ ($l_{\textrm{diffusion}}\ll l_{\textrm{penetration}}$), the particles can only explore a small portion of the potentials.  We call this the ``fast'' drive regime.
			
			$\bullet$  For RnT particles, the effective diffusivity over long time scales is given by $D=v^2/2\alpha$ and so we pick $l_{\textrm{diffusion}}=\sqrt{v^2/\alpha\omega}$.  Eq.\ (\ref{eq:rnt pde lab}) becomes
			\begin{equation}
				\frac{\partial\rho_{\pm}}{\partial\tilde{t}}=-\frac{\partial}{\partial\tilde{y}}\left[\left(\pm\gamma_{\omega}^{-1/2}-\epsilon\cos\tilde{t}-\gamma_f\gamma_{\omega}^{-1/2}\tilde{\Phi}'(\tilde{y})\right)\rho_{\pm}\right]-\gamma_{\omega}^{-1}\rho_{\pm}+\gamma_{\omega}^{-1}\rho_{\mp},
				\label{eq:rnt pde rest dim}
			\end{equation}
			where the parameters are defined as
			\begin{equation}
				\epsilon=\sqrt{\frac{a^2\omega\alpha}{v^2}},\ \ \ \gamma_{\omega}=\frac{\omega}{\alpha},\ \ \ \gamma_f=\frac{\mu f}{v}.
				\label{eq:def active params}
			\end{equation}
			Here, $\epsilon$ plays the same physical role as that of the passive particles and as such, we consider corrections to linear order in $\epsilon$.  The key difference between passive and active particles is the introduction of a new time scale $\alpha^{-1}$.  The parameter $\gamma_{\omega}$ controls the number of times a RnT particle tumbles during one cycle.  This introduces a new regime where the drive is faster than the tumbling ($\omega\gg\alpha$), which cannot happen for passive particles.  Finally, the parameter $\gamma_f$ compares the force the potential exerts on the particles to their propulsion force.  We take $\gamma_f<1$ so that the RnT particles are able to climb up the potentials.
			
			It should be noted that the quantity $\gamma_f\gamma_{\omega}^{-1/2}$ in Eq.\ (\ref{eq:rnt pde rest dim}) is analogous to $\gamma$ for passive particles.  This is easily seen by taking $D\sim v^2/\alpha$.  As we will see, this means that the ratio $l_{\textrm{diffusion}}/l_{\textrm{penetration}}$ will be important in determining the behavior of RnT particles, in addition to $\gamma_{\omega}$.
		
		\subsection{\label{subsec:Solution}Mechanical force and linear response function}
			In linear response theory, we may ask how much extra force $\Delta F$ the pistons exert on the suspension of particles due to a particular motion $a(t)$ of the pistons.  Mathematically, we write
			\begin{equation}
				\Delta F(t)=\int B(t-t')a(t')dt',
				\label{eq:B def}
			\end{equation}
			where $B(t)$ is the linear response function linking the perturbations to the responses.  In frequency space, Eq.\ (\ref{eq:B def}) can be written as $\Delta F_{\omega}=B_{\omega}a_{\omega}$, that is, a drive with frequency $\omega$ leads to a response with the same frequency in the linear regime.
			
			As discussed, we solve Eqs.\ (\ref{eq:diffusion pde rest dim}) and (\ref{eq:rnt pde rest dim}) to linear order in $\epsilon$.  The details of the calculation for all cases can be found in Appendix \ref{app:Solution Details}.  For all of the cases, the resulting density can be written in the form
			\begin{equation}
				\rho(y,t)=\rho^{(0)}(y)+2\epsilon\textrm{Re}\left\{p(y)e^{i\omega t}\right\},
			\end{equation} 
			where $\rho=\rho_++\rho_-$ for RnT particles and $p(y)$ describes the position dependence of the perturbation to the density.  The total mechanical force applied to the suspension can be computed as
			\begin{equation}
				F(t)=-\int_{-\infty}^{\infty}\Phi'(y)\rho(y,t)dy=F^{(0)}+\Delta F(t),
				\label{force}
			\end{equation}
			where $F^{(0)}$ is the force needed to keep the pistons stationary when there is no time-dependent drive.  For the ramp potential in 1D, this is the usual ideal gas pressure/force $F^{(0)}=-\rho_0D/\mu$, where $D=v^2/2\alpha$ for RnT particles.  For the V-shaped potential, $F^{(0)}=0$.  The extra force generated by the motion $a(t)=a\sin\omega t$ in the pistons is
			\begin{equation}
				\Delta F(t)=-2\epsilon\int_{-\infty}^{\infty}\Phi'(y)\re\left\{p(y)e^{i\omega t}\right\}dy.
				\label{eq:Delta F}
			\end{equation}
			Using the relation $\Delta F_{\omega}=B_{\omega}a_{\omega}$ and after some algebra (Appendix \ref{app:Response Function}), we arrive at
			\begin{equation}
				B_{\omega}=-\frac{2i}{l_{\textrm{diffusion}}}\int_{-\infty}^{\infty}\Phi'(y)p(y)dy,
				\label{response function}
			\end{equation}
			
			\begin{figure}
				\centering
				\includegraphics[scale=0.35]{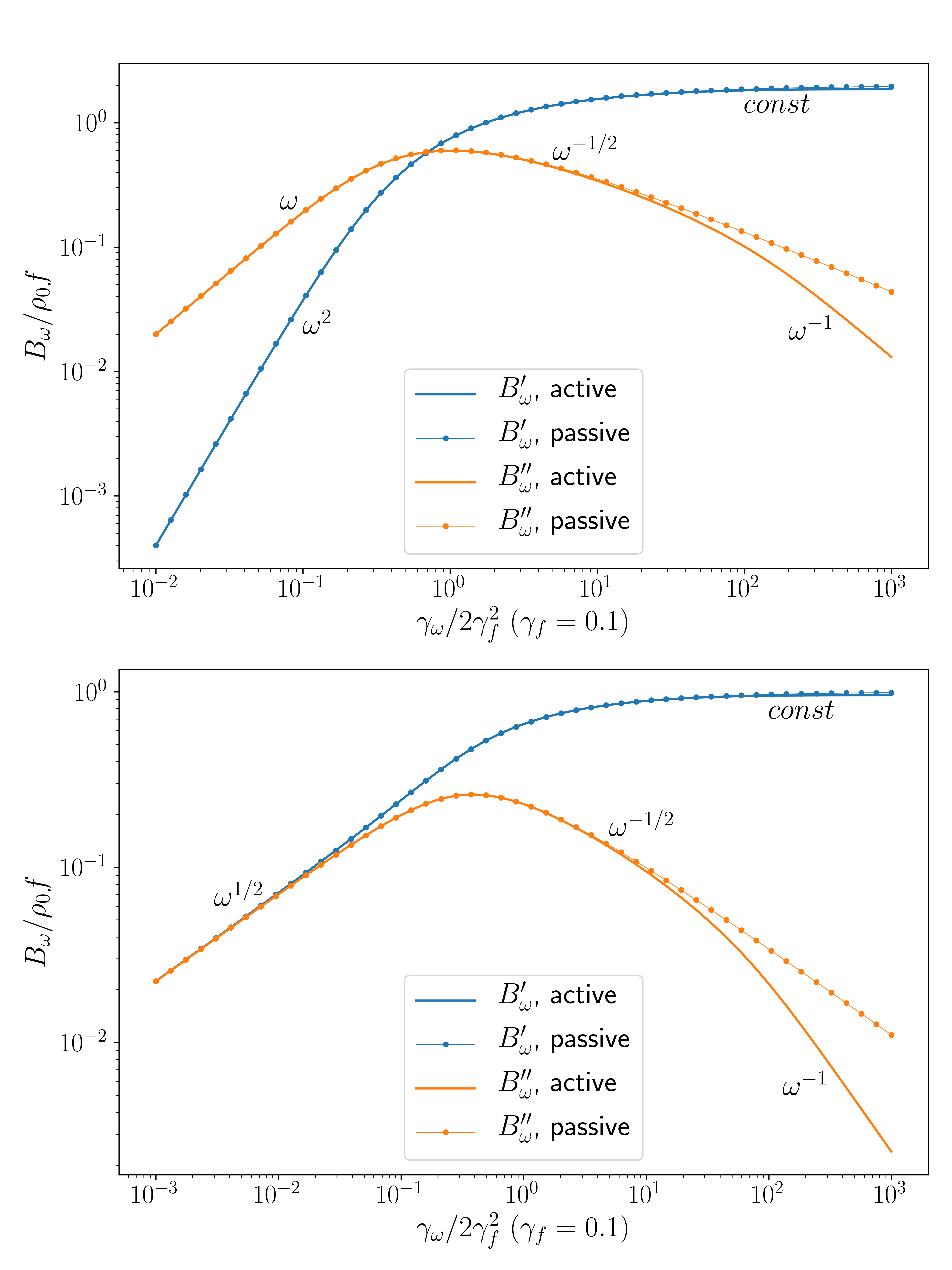}
				\caption{\textbf{Top:} Response function for V-shaped potential.  \textbf{Bottom:} Response function for ramp potential.  Real (blue) and imaginary (orange) parts of the response function for passive (dots) and RnT (solid line) particles.  For the rescaled frequency $\tilde{\omega}=\gamma_{\omega}/2\gamma_f^2$, we show $\gamma_f=0.1$ for the RnT particles.  There are three regimes: $\tilde{\omega}\ll1$, $1\ll\tilde{\omega}\ll\gamma_f^{-2}$, and $\tilde{\omega}\gg\gamma_f^{-2}$.}
				\label{fig:Bw passive active v ramp}
			\end{figure}
			$\bullet$  We start with the results for the \textbf{double pistons} (\textbf{V-shaped potential}).  For passive particles, the response function is
			\begin{equation}
				B_{\omega}=i\rho_0f\left(-2i-\gamma^2+\gamma\sqrt{\gamma^2+4i}\right),
			\end{equation}
			where $\gamma=\sqrt{\mu^2f^2/\omega D}$ (Eq.\ (\ref{eq:def passive params})).  For RnT particles, it is
			\begin{equation}
				B_{\omega}=i\rho_0f\left(-2i-\frac{2\gamma_f^2}{\gamma_{\omega}}+\frac{\sqrt{2}\gamma_f}{\gamma_{\omega}^{1/2}}\sqrt{\frac{2\gamma_f^2}{\gamma_{\omega}}+4i-2\gamma_{\omega}}\right),
			\end{equation}
			where $\gamma_{\omega}=\omega/\alpha$ and $\gamma_f=\mu f/v$ (Eq.\ (\ref{eq:def active params})).  
			
			Both cases can be compared in terms of a single rescaled frequency $\tilde{\omega}=\gamma_{\omega}/2\gamma_f^2=\gamma^{-2}$ (Fig.\ \ref{fig:Bw passive active v ramp}) by equating the diffusivities $v^2/2\alpha$ and $D$.  There are a total of three regimes.  For passive particles, the two regimes $\tilde{\omega}\ll1$ and $\tilde{\omega}\gg1$ correspond to the aforementioned slow and fast drives.  The key is that for RnT particles there is an additional region $\tilde{\omega}\gg\gamma_f^{-2}$ or $\omega\gg\alpha$, in which the pistons oscillate many times during a single tumble, that is, the particles appear persistent on the time scale of the drive.
			\begin{figure}
				\centering
				\includegraphics[scale=0.35]{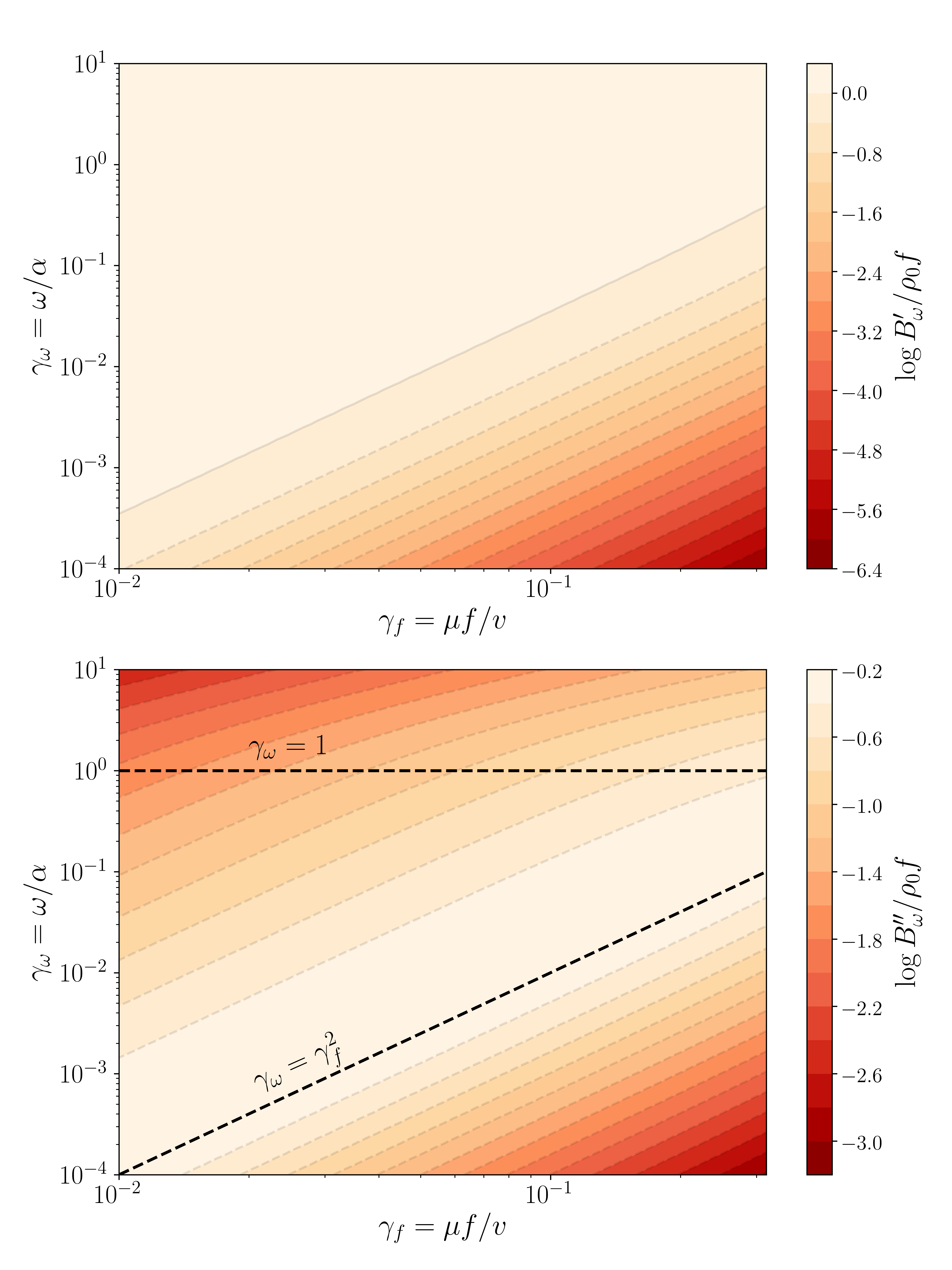}
				\caption{Level sets of $B_{\omega}$ for RnT particles in the V-shaped potential.  The contours correspond to fixed $\tilde{\omega}$.  \textbf{Top:} Real part. \textbf{Bottom:} Imaginary part.  There are two crossovers indicated by the black dashed lines: $\gamma_{\omega}\sim\gamma_f^2$, which divides the slow and fast regimes; and $\gamma_{\omega}\sim1$, which divides the diffusive and persistent regimes.}
				\label{fig:Bw contour v}
			\end{figure}
			The region $\omega\ll\alpha$ where the passive and active particles have similar behaviors is the passive limit of RnT particles, which can be obtained by taking $v,\alpha\rightarrow\infty$ while holding $v^2/2\alpha$ constant, or equivalently moving along the contours $\gamma_{\omega}=2\tilde{\omega}\gamma_f^2$ for fixed $\tilde{\omega}$ (Fig.\ \ref{fig:Bw contour v}).
			
			$\bullet$  For the \textbf{single piston} (\textbf{ramp potential}), the response function for passive particles is
			\begin{equation}
				B_{\omega}=\frac{i^{3/2}\rho_0f}{2}\left(\gamma-\sqrt{\gamma^2+4i}\right).
			\end{equation}
			For RnT particles, the expression is rather cumbersome and is presented in Appendix \ref{app:Variable Definitions} (Eq.\ (\ref{eq:response ramp potential})).  We see the same three regions as with the V-shaped potential.  The key difference is the frequency dependence of the response functions in the slow drive regime $\tilde{\omega}\ll1$ (Fig.\ \ref{fig:Bw passive active v ramp}, right).
		
	\section{\label{sec:Discussion}Discussion}		
		To briefly summarize, we obtained the linear response functions of a system of passive particles (diffusivity $D$) or run-and-tumble particles (swim speed $v$ and tumble rate $\alpha$) under external forcing by considering a suspension of these particles driven by a single piston or between two pistons, which we represented by linear potentials.  In particular, we considered a periodic drive $a(t)=a\sin\omega t$, from which we extracted the frequency dependence of the response of the mechanical force to the drive.
		
		The real and imaginary parts of the response function $B_{\omega}=B_{\omega}'+iB_{\omega}''$ are associated with the storage and dissipation of energy, respectively.  Note that the excess mechanical force $\Delta F$ and the rate at which excess work $\Delta\dot{W}=\Delta F\dot{a}$ is performed on the particles can be written as
		\begin{equation}
			\Delta F(t)=aB_{\omega}'\sin\omega t+aB_{\omega}''\cos\omega t
		\end{equation}
		and
		\begin{equation}
			\Delta\dot{W}=\frac{1}{2}a^2\omega B_{\omega}'\sin2\omega t+\frac{1}{2}a^2\omega B_{\omega}''\left(1+\cos2\omega t\right),
			\label{eq:work rate}
		\end{equation}
		where the first terms correspond to in-phase responses and the second terms, out-of-phase responses.

		The work performed on the system by the drive is stored and dissipated by the particles.  The first term on the right-hand side of Eq.\ (\ref{eq:work rate}), which is due to the in-phase ``elastic"-like response, is related to storage.  In this non-interacting ideal gas-like system, energy is stored when the particles climb up and spend time in the potentials.  Even though no energy is stored on average over time, we identify the amplitude of the rate of storage as
		\begin{equation}
		\Delta\dot{U}\approx\frac{1}{2}a^2\omega B_{\omega}'.
		\label{eq:storage rate}
		\end{equation}
		The second term on the right-hand side of Eq.\ (\ref{eq:work rate}), which comes from the out-of-phase ``viscous"-like response, is related to dissipation.  As expected, this term has a non-zero average over each cycle since the dissipated energy cannot be returned.  Thus the average rate of dissipation is 
		\begin{equation}
			\langle\dot{Q}\rangle=\frac{1}{2}a^2\omega B_{\omega}''.
			\label{eq:dissipation rate}
		\end{equation}
		
		There are three distinct regimes of response, depending on the driving frequency $\omega$:
		\begin{enumerate}[label = (\roman*)]
			\item Slow drive---In this regime, we have $\sqrt{D/\omega}\gg D/\mu f$ (for the case of active particles, $D=v^2/2\alpha$), that is, the region over which particles diffuse in a cycle is much greater than the penetration depth of the linear potentials.  As a consequence, the particles can explore the confining potentials and effectively equilibrate with them.  
			\item Fast drive, diffusive particles---In this regime, we instead have $\sqrt{D/\omega}\ll D/\mu f$.  In other words, the particles will not have sufficient time to diffuse sufficiently far in a cycle to explore the confinement and thus will not equilibrate with the potentials.
			\item Fast drive, persistent particles---In this regime, the drive is sufficiently fast $\omega\gg\alpha$ such that the rate of driving is faster than the tumble rate of the particles and the particles appear persistent on the time scale of the drive.
		\end{enumerate}
		Note that in the first two regimes (Fig.\ \ref{fig:Bw passive active v ramp}), the passive and active particles have similar behaviors.  This is the passive limit for RnT particles, that is, for sufficiently slow drive, the RnT particles effectively behave as passive particles with diffusivity $v^2/2\alpha$.  As such, for these two regimes, we will consider particles with diffusivity $D$, keeping in mind that we can simply replace $D$ with $v^2/2\alpha$ for active particles.  
		
		We now illustrate out results with scaling arguments.
		
		\begin{table}
			\centering
			\begin{tabular}{| c | c | c | c | c |}
				\cline{2-5}
				\multicolumn{1}{c |}{} & \multicolumn{2}{c |}{\textbf{V-shaped potential}} & \multicolumn{2}{c |}{\textbf{ramp potential}}\\
				\multicolumn{1}{c |}{} & $\langle\dot{Q}\rangle$ & $\Delta\dot{U}$ & $\langle\dot{Q}\rangle$ & $\Delta\dot{U}$\\\hline
				slow drive & $\rho_0\frac{D}{\mu f}\frac{(a\omega)^2}{\mu}$ & $\rho_0\frac{D}{\mu f}\frac{(a\omega)^2}{\mu}\frac{1}{\gamma^2}$ & $\rho_0\sqrt{\frac{D}{\omega}}\frac{(a\omega)^2}{\mu}$ & $\rho_0\sqrt{\frac{D}{\omega}}\frac{(a\omega)^2}{\mu}$\\
				($\tilde{\omega}\ll1$) & & & & \\\hline
				fast drive, & & & & \\ 
				diffusive particles & $\rho_0\sqrt{\frac{D}{\omega}}\frac{(a\omega\gamma)^2}{\mu}$ & $\rho_0a^2f\omega$ & $\rho_0\sqrt{\frac{D}{\omega}}\frac{(a\omega\gamma)^2}{\mu}$ & $\rho_0a^2f\omega$\\
				($1\ll\tilde{\omega}\ll\gamma_f^{-2}$) & & & & \\\hline
				fast drive, & & & & \\
				persistent particles & $\rho_0\frac{a^2\mu f^2\alpha}{v}$ & $\rho_0a^2f\omega$ & $\rho_0\frac{a^2\mu f^2\alpha}{v}$ & $\rho_0a^2f\omega$\\
				($\gamma_f^{-2}\ll\tilde{\omega}$) & & & & \\\hline
			\end{tabular}
			\caption{Average rate of dissipation $\langle\dot{Q}\rangle$ and the characteristic rate of storage $\Delta\dot{U}$ for a drive $a(t)=a\sin\omega t$.  As defined in the text, $\tilde{\omega}=\omega D/\mu^2f^2$.  Physically, the first regime corresponds to $l_{\textrm{diffusion}}\gg l_{\textrm{penetration}}$; the second regime to $l_{\textrm{diffusion}}\ll l_{\textrm{penetration}}$ and $\omega\ll\alpha$; and the third regime to $\omega\gg\alpha$.  Note that for slow and fast drives, the case of active particles can be obtained by replacing $D$ with $v^2/2\alpha$.}
			\label{tab:dissipation storage}
		\end{table}
		\subsection{\label{subsec:slow drive}Slow drive}
			When the drive is sufficiently slow, the drive probes time scales greater than the relaxation time of diffusing particles in the linear potentials and we have $l_{\textrm{diffusion}}\gg l_{\textrm{penetration}}$ ($\gamma\gg1$).  Note that in this regime, the system can be described by an adiabatic approximation.  We start with the average rate at which particles dissipate work performed by the drive (Eq.\ (\ref{eq:dissipation rate})):
			\begin{equation}
				\langle\dot{Q}\rangle_{\textrm{V-shaped}}\sim\rho_0\frac{D}{\mu f}\frac{(a\omega)^2}{\mu},
				\label{eq:disspation, v-shaped, slow}
			\end{equation}
			\begin{equation}
				\langle\dot{Q}\rangle_{\textrm{ramp}}\sim\rho_0\sqrt{\frac{D}{\omega}}\frac{(a\omega)^2}{\mu}.
				\label{eq:dissipation, ramp, slow}
			\end{equation}
			Before we continue, it is useful to interpret the dissipation rates for both the slow and fast drive regimes in terms of the particle current generated by the drive (Appendix \ref{app:Currents}).  The current can be written as $J_x(y,t)=\rho(y,t)v_x(y,t)$, where $v_x$ is the average drift velocity of the particles relative to the static fluid.  The dissipation rate due to drag can then be computed as
			\begin{equation}
				\dot{Q}=\int_{-\infty}^{\infty}\rho\frac{v_x^2}{\mu}dy.
			\end{equation}
			Upon time averaging, we finally note that the average dissipation rate scales as
			\begin{equation}
				\langle\dot{Q}\rangle\sim\rho_0L\frac{V^2}{\mu},
				\label{eq:dissipation form}
			\end{equation}
			where $L$ is the characteristic decay length of the current, or $\rho_0L$ is the number of particles contributing to dissipation, and $V$ is the characteristic drift velocity of those particles.
			
			For the \textbf{double pistons} (\textbf{V-shaped potential}), we find in terms of Eq.\ (\ref{eq:dissipation form}) $L\sim D/\mu f$, the penetration depth, and $V\sim a\omega$, the characteristic speed of the pistons.  Since the particles can equilibrate with the potential, they on average drift with the same velocity as the pistons.  Thus, we have the net transport of $\rho_0D/\mu f$ particles (the total number of trapped particles) each with a dissipation rate of $(a\omega)^2/\mu$ (Eq.\ (\ref{eq:disspation, v-shaped, slow})).
			
			For the \textbf{single piston} (\textbf{ramp potential}), particles are still transported with the piston.  However, the key difference from the V-shaped potential is in the number of particles that contribute to dissipation.  This is due to the presence of a bulk.  Again using the language of Eq.\ (\ref{eq:dissipation form}), we have $L\sim\sqrt{D/\omega}$, the diffusion distance.  In the bulk, only particles within a diffusion distance of the piston will equilibrate with the potential.  Particles farther than that remain unaffected by the movement of the pistons during a cycle.  Thus, unlike with the V-shaped potential, $\rho_0\sqrt{D/\omega}$ particles are instead transported each with dissipation rate $(a\omega)^2/\mu$ (Eq.\ (\ref{eq:dissipation, ramp, slow})).
			
			The storage rates given by Eq.\ (\ref{eq:storage rate}) are
			\begin{equation}
				\Delta\dot{U}_{\textrm{V-shaped}}\sim\rho_0\frac{D}{\mu f}\frac{(a\omega)^2}{\mu}\frac{1}{\gamma^2}\ll\langle\dot{Q}\rangle_{\textrm{V-shaped}},
				\label{eq:storage, v-shaped, slow}
			\end{equation}
			\begin{equation}
				\Delta\dot{U}_{\textrm{ramp}}\sim\rho_0\sqrt{\frac{D}{\omega}}\frac{(a\omega)^2}{\mu}\sim\langle\dot{Q}\rangle_{\textrm{ramp}}.
				\label{eq:storage, ramp, slow}
			\end{equation}
			For the \textbf{double pistons} (\textbf{V-shaped potential}), we expect that very little energy will be stored during a cycle.  Indeed, we find that the storage rate is significantly smaller than the dissipation rate since $\gamma\gg1$.  This is due to the symmetry of the potential.  When the potential shifts positions slowly, depending on the direction, the potential energy on one side of the origin increases slightly from an influx of particles while the potential energy on the other side decreases by roughly an equal amount from an efflux of particles.  
			
			For the \textbf{single piston} (\textbf{ramp potential}), the situation is different since particles in the bulk do not have any potential energy; a flow in or out of the potential region will lead to a much more significant change in the potential energy of the system.  In fact, we find that the storage rate is of order the dissipation rate.  As the piston moves towards the bulk, the influx of particles into the potential region leads to an increase in potential energy while the efflux of particles from the bulk does nothing.  Since the drive is slow, these particles have sufficient time to leave the region and dissipate any energy given to them.
			
			Note that since the slow drive regime can be described by an adiabatic approximation, these results should hold for any confining potential.
			
		\begin{figure}
			\centering
			\includegraphics[scale=0.35]{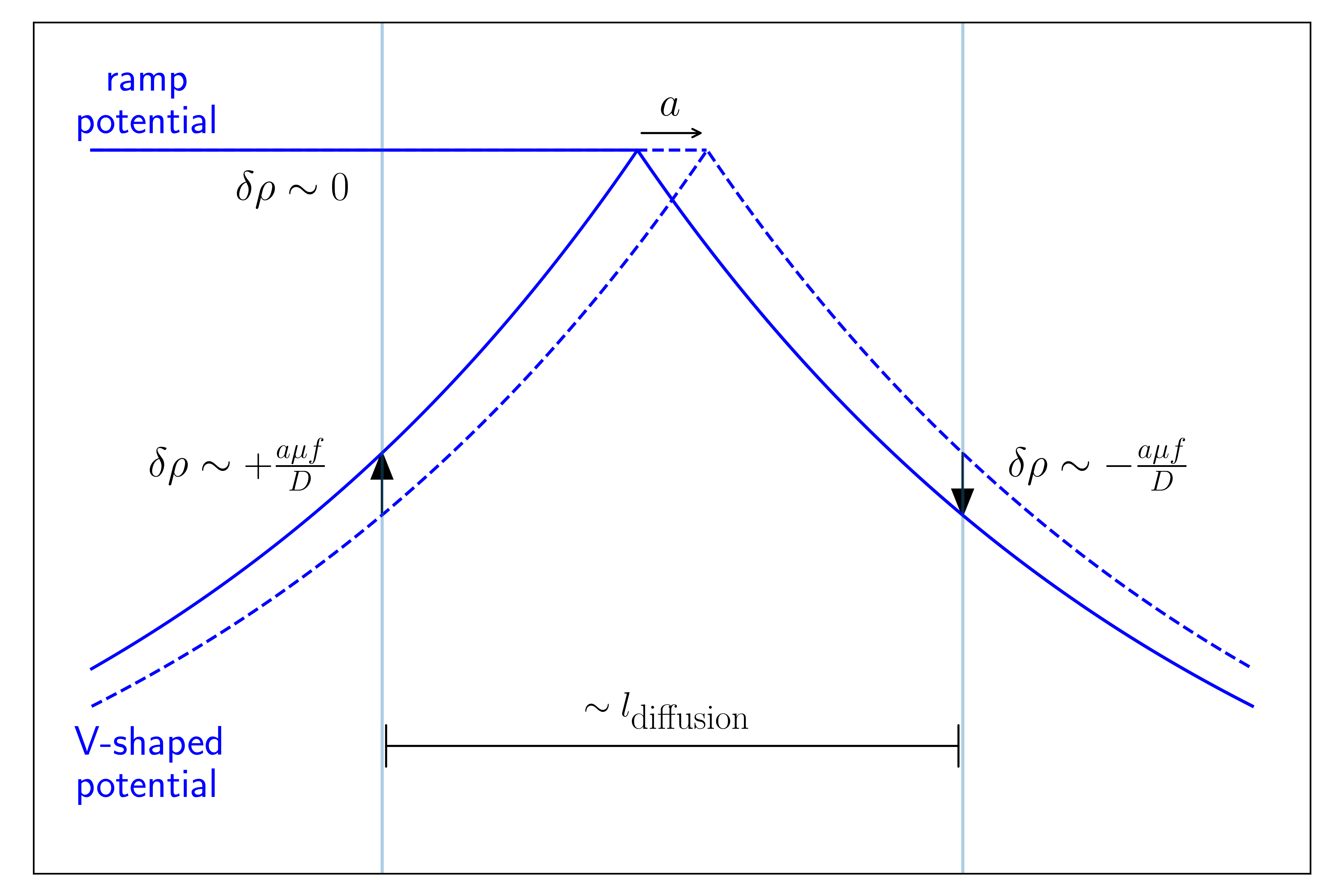}
			\caption{Schematic of the density gradients that give rise to dissipation through diffusion for fast drive.  If the drive were sufficiently slow, the original density (solid blue) would relax to the new shifted density (dashed blue).  However, for fast drive, only particles within $l_{\textrm{diffusion}}$ of the origin will see a change in the environment and begin to relax through diffusion.}
			\label{fig:density gradients}
		\end{figure}
		\subsection{\label{subsec:Fast drive}Fast drive, diffusive particles}
			When the pistons are driven quickly, the particles do not have sufficient time to relax in the potential regions since $l_{\textrm{diffusion}}\ll l_{\textrm{penetration}}$ ($\gamma\ll1$).  In this case, the dissipation rates (Eq.\ (\ref{eq:dissipation rate})) go as
			\begin{equation}
				\langle\dot{Q}\rangle_{\textrm{V-shaped}}\sim\langle\dot{Q}\rangle_{\textrm{ramp}}\sim\rho_0\sqrt{\frac{D}{\omega}}\frac{(a\omega\gamma)^2}{\mu}.
				\label{eq:dissipation, passive fast}
			\end{equation}
			Here for both potentials, we find $L\sim\sqrt{D/\omega}$ and $V\sim a\omega\gamma$, which is much slower than the drive.  It is not too surprising that the decay length scale of the current is the diffusion distance.  For fast drive, particles farther than that distance from the origin will not see a change in their environment (they experience the same constant force $\pm f$ throughout each cycle) and thus the density in those regions will roughly remain unaffected by the drive.  However, particles within a diffusion distance of the origin will notice the shift in the potentials and begin to relax to a new steady-state density (Fig.\ \ref{fig:density gradients}).  This relaxation of density near the origin generates the current.  
			
			To understand how the particular speed $V\sim a\omega\gamma$ arises, note that the condition $l_{\textrm{diffusion}}\ll l_{\textrm{penetration}}$ can be written as $fl_{\textrm{diffusion}}\ll D/\mu$ ($\equiv k_BT_{\textrm{eff}}$).  Therefore, the effect of the potentials is weak compared to that of diffusion, that is, the diffusive flux due to density gradients dominates over the advective flux due to the potentials.  The density gradients of interest are those that occur over the region of size $\sim l_{\textrm{diffusion}}$ around the origin since only the density in that region evolves.  In other words, by shifting the potential a distance $a$ on time-scale $\omega^{-1}$, the density at the boundaries of the region bounded by $\pm l_{\textrm{diffusion}}$ is either increased or decreased by a fixed amount $\delta\rho$ (Fig.\ \ref{fig:density gradients}).  The unperturbed density is given by $\rho(x)=\rho_0e^{-\mu f|x|/D}$.  Shifting and taking the difference, we find for small $a$
			\begin{equation}
				\delta\rho=\rho(\pm l_{\textrm{diffusion}})-\rho(\pm l_{\textrm{diffusion}}-a)\approx\mp\frac{\rho_0a\mu f}{D}.
			\end{equation}
			Therefore, the diffusive current over this region can be computed as
			\begin{equation}
				J\sim D\frac{\delta\rho}{l_{\textrm{diffusion}}}\sim\rho_0a\omega\gamma,
			\end{equation}
			from which we obtain $V\sim a\omega\gamma$.  Note that although $\delta\rho\approx0$ at $-l_{\textrm{diffusion}}$ for the case of the ramp potential (the bulk does not change very much), the density gradient still scales the same and we would still obtain the same characteristic drift speed.
			
			The storage rates (Eq.\ (\ref{eq:storage rate})) go as
			\begin{equation}
				\Delta\dot{U}_{\textrm{V-shaped}}\sim\Delta\dot{U}_{\textrm{ramp}}\sim(\rho_0a)(fa)\omega.
				\label{eq:storage, passive fast}
			\end{equation}
			Note that since particles on average transport much slower than the pistons, we can treat them on average as stationary over a period.  Thus, when a piston moves a distance $a$ towards particles, $\rho_0a$ particles are forced a distance of order $a$ into the potential region.  The potential energy increase per particle is then roughly $fa$.  This happens at a rate $\omega$ and so we obtain the same scaling.
		
		\subsection{\label{subsec:active fast drive}Fast drive, persistent particles}
			As we saw in the previous two regimes, there is little difference between the passive and active particles aside from a re-defined diffusion constant.  This is due to the diffusive nature of the active particles on long time-scales.  In this regime, however, the drive rate is larger than the tumbling rate ($\omega\gg\alpha$) and the active particles will be persistent on the time-scales of the drive.  The dissipation rates (Eq.\ (\ref{eq:dissipation rate})) are given by
			\begin{equation}
				\langle\dot{Q}\rangle_{\textrm{V-shaped}}\sim\langle\dot{Q}\rangle_{\textrm{ramp}}\sim\rho_0\frac{a^2\mu f^2\alpha}{v}.
				\label{eq:persistent dissipation}
			\end{equation}
			If we compute the currents as we did for the slow and fast drive regimes for diffusive particles, we find that the dissipation through particle currents is
			\begin{equation}
				\langle\dot{Q}\rangle\sim\rho_0\frac{a^2\mu f^2\alpha}{v}\left(\frac{\omega}{\alpha}\right)^2,
			\end{equation}
			which is different from the work the drive performs on the system.  To illustrate the source of this discrepancy, consider a passive Brownian particle and an active swimmer, both acted on by an external force $f_{\textmd{ext}}$.  For the Brownian particle, we know that in addition to diffusion, it will on average drift with a speed $v=\mu f_{\textrm{ext}}$.  The average work rate of the external force is $\dot{w}_{\textrm{ext}}=f_{\textrm{ext}}v=\mu f_{\textrm{ext}}^2=\dot{q}$, which is the rate of dissipation through friction.  For a swimmer, on a time-scale shorter than the correlation time $\alpha^{-1}$, the drift velocity is $v=\mu(f_p+f_{\textrm{ext}})$, where $f_p$ is the propulsion force, and the rate of dissipation is $\dot{q}=(f_p+f_{\textrm{ext}})v=\mu(f_p+f_{\textrm{ext}})^2$.  The change in the rate of dissipation from when there is no external force is $\delta\dot{q}=\mu(f_p+f_{\textrm{ext}})^2-\mu f_p^2=2\mu f_pf_{\textrm{ext}}+\mu f_{\textrm{ext}}^2$.  Naively, we would attribute this extra dissipation to work performed by the external force; however, a quick calculation reveals that the work rate of the external force is given by $\dot{w}_{\textrm{ext}}=f_{\textrm{ext}}v=\mu f_{\textrm{ext}}(f_p+f_{\textrm{ext}})\ne\dot{q}$.  Note that the work rate performed by the swimmer propulsion is $\dot{w}_p=\mu f_p(f_p+f_{\textrm{ext}})$.  Therefore, the dissipation for a swimmer not only accounts for the work performed by the external force, but also the effect of that external force on the motoring activity of the swimmer, that is, the external force changes the work rate performed by self-propulsion.
			
			We should note that different physical situations are possible with regards to the interplay between external drive and internal motoring activity of the swimmers.  We assumed above---as was natural in the context of our work---that the propulsion force $f_p$ remains unaffected by the external drive.  In other systems, it is possible, for instance, that the power output by a propelled motor, $f_pv$, is a constant, in which case the propulsion force itself will have to be dependent on the external drive (since velocity depends on it).  Various intermediate cases, between constant propulsion force and constant power output, are also possible.  We do not analyze all these possibilities in this paper.
			
			The storage rates (Eq.\ (\ref{eq:storage rate})) are given by
			\begin{equation}
			\Delta\dot{U}_{\textrm{V-shaped}}\sim\Delta\dot{U}_{\textrm{ramp}}\sim(\rho_0a)(fa)\omega.
			\label{eq:storage, active fast}
			\end{equation}
			Not surprisingly, this scaling is the same as fast drive since the same reasoning applies, that is, moving a piston a distance $a$ forces $\rho_0a$ particles into the potential region and gives them a potential energy $fa$ on a time scale $\omega$.  This is independent of whether these particles are passive or active.
	
	\section{\label{sec:Conclusion}Concluding remarks}
		To conclude, we examined the linear response of a system of passive or active (run-and-tumble) particles to a time-periodic drive.  We found that the active suspension responds in a way similar to that of passive particles when the frequency of drive is smaller than the inverse persistence time of the active particles.  At higher frequencies (larger than the inverse persistence time), however, the persistence of the active particles changes the response a great deal; in particular, we found that the dissipation rate of the active particles changes not only due to work performed by the drive but also due to the effect of the drive on the motoring activity of the particles, that is, the drive changes the work rate of the particles own self-propulsion.
		
		The phenomena we studied here has an interesting analogy with a very old problem examined first by J.\ Fourier himself in his treatise from where Fourier transforms originate \cite{Fourier French,Fourier English}.  In particular, Fourier considered the diffusion of heat through a medium due to temperature variations at the boundary; for example, the heating and cooling below ground as a result of temperature variations throughout the seasons.  In our notation, he found that the temperature variations as a function of the depth went as
		\begin{equation}
			\delta T(y,t)=\delta T_0e^{\sqrt{i\frac{\omega}{D}}y+i\omega t}.
		\end{equation}
		The key feature is that the wavelength of the variations is the same as the decay length.  For us, for example for the ramp potential (Eq.\ (\ref{eq:ramp solution})) in the bulk, the density variations due to a drive effectively at the boundaries (since the drive amplitude is much smaller than how far the particles diffuse in a single period, $l_{\textrm{diffusion}}$) have the same form
		\begin{equation}
			\delta\rho(y,t)=\delta\rho_0e^{\sqrt{i\frac{\omega}{D}}y+i\omega t}.
		\end{equation}
		It is interesting to note that for active particles in the persistent limit (drive frequency larger than the inverse persistence time), the density variations (Eq.\ (\ref{eq:rnt, ramp, y<0})) become
		\begin{equation}
			\delta\rho(y,t)=\delta\rho_0e^{\frac{\alpha}{v}y}e^{i\left(-\frac{\omega}{v}y+\omega t\right)}.
		\end{equation}
		where now the wavelength of the variations (swim distance in a cycle) is much shorter than the decay length (persistence length of the active particles).
	
	\section*{Acknowledgments}
		This work was supported primarily by the MRSEC Program of the National Science Foundation under Award Number DMR-1420073.  We thank J.-F.\ Joanny and W.\ Srinin for stimulating discussions and their insightful comments.  
		
		We were fortunate to have worked in the same department with Pierre Hohenberg and one of us (AYG) had the opportunity to discuss with him this work at its early stages.  We therefore feel honored to submit this paper to the journal dedicated to his memory.
	
	\appendix
	\section{\label{app:Solution Details}Details of solution}
		Here, we show the details of obtaining the solutions in Sec.\ \ref{sec:Calculation} for passive or RnT particles in the V-shaped and ramp potentials.
		
		\subsection{\label{subapp:passive, v}Passive particles, V-shaped potential}
			In the frame of the potential, the Fokker-Planck equation is
			\begin{equation}
				\frac{\partial\rho}{\partial t}=\frac{\partial}{\partial y}\Big[\left(a\omega\cos\omega t+\mu\sgn(y)\right)\rho\Big]+D\frac{\partial^2\rho}{\partial y^2},
			\end{equation}
			where $y=x-a\sin\omega t$.  Picking the time and length scales to be $\omega^{-1}$ (time-scale of the drive) and $\sqrt{D/\omega}$ (diffusion distance), we arrive at (Eq.\ (\ref{eq:diffusion pde rest dim}))
			\begin{equation}
				\frac{\partial\rho}{\partial\tilde{t}}=\frac{\partial}{\partial\tilde{y}}\Big[\left(\epsilon\cos\tilde{t}+\gamma\sgn(\tilde{y})\right)\rho\Big]+\frac{\partial^2\rho}{\partial\tilde{y}^2}
			\end{equation}
			
			The transient solution can be written as $\rho(\tilde{y},\tilde{t})=\rho^{(0)}(\tilde{y})+\epsilon\rho^{(1)}(\tilde{y},\tilde{t})+O(\epsilon^2)$.  To zeroth order, we have
			\begin{equation}
				0=\frac{\partial}{\partial\tilde{y}}\left[\gamma\sgn(\tilde{y})\rho^{(0)}+\frac{\partial\rho^{(0)}}{\partial\tilde{y}}\right],
			\end{equation}
			which gives the usual exponential distribution
			\begin{equation}
				\rho^{(0)}(\tilde{y})=\rho_0e^{-\gamma|\tilde{y}|}.
			\end{equation}
			
			The first order correction can be written as $\rho^{(1)}(\tilde{y},\tilde{t})=p(\tilde{y})e^{i\tilde{t}}+p^*(\tilde{y})e^{-i\tilde{t}}$, where $p^*$ is the complex conjugate of $p$.  For $\tilde{y}<0$, $p(\tilde{y})$ satisfies
			\begin{equation}
				\frac{d^2p}{d\tilde{y}^2}-\gamma\frac{dp}{d\tilde{y}}-ip=-\frac{\rho_0}{2}\gamma e^{\gamma\tilde{y}}.
			\end{equation}
			Requiring that $\lim\limits_{\tilde{y}\rightarrow-\infty}\rho=0$, we obtain
			\begin{equation}
				p(\tilde{y}<0)=a_+e^{\xi\tilde{y}}+ce^{\gamma\tilde{y}},
			\end{equation}
			where $\xi=\frac{1}{2}\left(\gamma+\sqrt{\gamma^2+4i}\right)$ and $c=-i\rho_0\gamma/2$.  For $\tilde{y}>0$, taking $\gamma\rightarrow-\gamma$ and requiring $\lim\limits_{\tilde{y}\rightarrow\infty}\rho=0$ gives
			\begin{equation}
				p(\tilde{y}>0)=b_-e^{-\xi\tilde{y}}+de^{-\gamma\tilde{y}},
			\end{equation}
			where $d=i\rho_0\gamma/2$.  Continuity in density and current at $\tilde{y}=0$ gives the two conditions
			\begin{align}
				a_++c&=b_-+d,\\
				(\xi-\gamma)a_+&=(-\xi+\gamma)b_-,
			\end{align}
			from which we find
			\begin{equation}
				a_+=-b_-=i\frac{\rho_0\gamma}{2}.
			\end{equation}
			Therefore, the first order correction can be written as
			\begin{equation}
				\rho^{(1)}(\tilde{y},\tilde{t})=-\rho_0\gamma\sgn(\tilde{y})\re\left\{i\left(e^{-\xi|\tilde{y}|}-e^{-\gamma|\tilde{y}|}\right)e^{i\tilde{t}}\right\}.
			\end{equation}
			
		\subsection{\label{subapp:passive, ramp}Passive particles, ramp potential}
			For the ramp potential $\Phi'(y)=f\theta(y)$, where $\theta$ is a step function.  Eq.\ (\ref{eq:diffusion pde rest dim}) is
			\begin{equation}
				\frac{\partial\rho}{\partial\tilde{t}}=\frac{\partial}{\partial\tilde{y}}\Big[\left(\epsilon\cos\tilde{t}+\gamma\theta(\tilde{y})\right)\rho\Big]+\frac{\partial^2\rho}{\partial\tilde{y}^2}.
			\end{equation}
			As before, we take the transient solution to be $\rho=\rho^{(0)}+\epsilon\rho^{(1)}+O(\epsilon^2)$.  To zeroth order, 
			\begin{equation}
				0=\frac{\partial}{\partial\tilde{y}}\left[\gamma\theta(\tilde{y})\rho^{(0)}+\frac{\partial\rho^{(0)}}{\partial\tilde{y}}\right],
			\end{equation}
			and assuming that $\lim\limits_{\tilde{y}\rightarrow-\infty}=\rho_0$, we have
			\begin{equation}
				\rho^{(0)}(\tilde{y})=\rho_0e^{-\gamma\theta(\tilde{y})\tilde{y}}.
			\end{equation}
			Again, taking $\rho^{(1)}=pe^{i\tilde{t}}+p^*e^{-i\tilde{t}}$, we get for $\tilde{y}<0$
			\begin{equation}
				\frac{d^2p}{d\tilde{y}^2}-ip=0,
			\end{equation}
			which gives
			\begin{equation}
				p(\tilde{y})=a_+e^{\sqrt{i}\tilde{y}}.
			\end{equation}
			For $\tilde{y}>0$,
			\begin{equation}
				\frac{d^2p}{d\tilde{y}^2}+\gamma\frac{dp}{d\tilde{y}}-ip=\frac{\rho_0\gamma}{2}e^{-\gamma\tilde{y}},
			\end{equation}
			the solution of which is
			\begin{equation}
				p(\tilde{y})=b_-e^{-\xi\tilde{y}}+de^{-\gamma\tilde{y}},
			\end{equation}
			where $\xi=\frac{1}{2}\left(\gamma+\sqrt{\gamma^2+4i}\right)$ and $d=i\rho_0\gamma/2$.  At $\tilde{y}=0$, continuity in density and current gives
			\begin{align}
				a_+&=b_-+d,\\
				\sqrt{i}a_+&=(\gamma-\xi)b_-,
			\end{align}
			and so
			\begin{equation}
				a_+=\frac{\xi-\gamma}{\xi-\gamma+\sqrt{i}}d,
			\end{equation}
			\begin{equation}
				b_-=-\frac{\sqrt{i}}{\xi-\gamma+\sqrt{i}}d.
			\end{equation}
			Thus,
			\begin{equation}
				\rho^{(1)}(\tilde{y},\tilde{t})=\rho_0\gamma\re
				\begin{cases}
					\frac{i(\xi-\gamma)}{\xi-\gamma+\sqrt{i}}e^{\sqrt{i}\tilde{y}}e^{i\tilde{t}},& \tilde{y}<0\\
					i\left(-\frac{\sqrt{i}}{\xi-\gamma+\sqrt{i}}e^{-\xi\tilde{y}}+e^{-\gamma\tilde{y}}\right)e^{i\tilde{t}},& \tilde{y}>0
				\end{cases}
				\label{eq:ramp solution}
			\end{equation}
		
		\subsection{\label{subapp:rnt, v}RnT particles, V-shaped potential}
			In this case, the Fokker-Planck equation is
			\begin{equation}
				\frac{\partial\rho_{\pm}}{\partial t}=-\frac{\partial}{\partial y}\Big[\left(\pm v-a\omega\cos\omega t-\mu f\sgn(y)\right)\rho_{\pm}\Big]-\alpha\rho_{\pm}+\alpha\rho_{\mp},
			\end{equation}
			where $y=x-a\sin\omega t$.  Picking the same time scales and length scales, $1/\omega$ (time scale of drive) and $\sqrt{v^2/\omega\alpha}$ (diffusion distance), we arrive at (Eq.\ (\ref{eq:rnt pde rest dim}))  
			\begin{equation}
				\frac{\partial\rho_{\pm}}{\partial\tilde{t}}=-\frac{\partial}{\partial\tilde{y}}\left[\left(\pm\gamma_{\omega}^{-1/2}-\epsilon\cos\tilde{t}-\gamma_f\gamma_{\omega}^{-1/2}\sgn(\tilde{y})\right)\rho_{\pm}\right]-\gamma_{\omega}^{-1}\rho_{\pm}+\gamma_{\omega}^{-1}\rho_{\mp}.
			\end{equation}
			
			The transient solution can be written as $\rho_{\pm}(\tilde{y},\tilde{t})=\rho_{\pm}^{(0)}(\tilde{y})+\epsilon\rho_{\pm}^{(1)}(\tilde{y},\tilde{t})+O(\epsilon^2)$.  The zeroth order solution satisfies
			\begin{equation}
				\frac{d}{d\tilde{y}}\left[(\pm1-\gamma_f\sgn(\tilde{y}))\rho_{\pm}^{(0)}\right]=-\gamma_{\omega}^{-1/2}\rho_{\pm}^{(0)}+\gamma_{\omega}^{-1/2}\rho_{\mp}^{(0)}.
				\label{eq:0th order eq}
			\end{equation}
			For $\tilde{y}<0$,
			\begin{equation}
				\frac{d}{d\tilde{y}}\pmat{\rho_+^{(0)}\\\rho_-^{(0)}}{1.2}=\gamma_{\omega}^{-1/2}\bmat{-\delta_+&\delta_+\\-\delta_-&\delta_-}{1.2}\pmat{\rho_+^{(0)}\\\rho_-^{(0)}}{1.2},
			\end{equation}
			where $\delta_{\pm}=1/(1\pm\gamma_f)$.  Requiring that $\lim\limits_{\tilde{y}\rightarrow-\infty}\rho=0$, we obtain
			\begin{equation}
				\pmat{\rho_+^{(0)}\\\rho_-^{(0)}}{1.2}=a\pmat{\delta_+\\\delta_-}{1.2}e^{\xi_0\tilde{y}},
			\end{equation}
			where $\xi_0=2\gamma_f\gamma_{\omega}^{-1/2}/(1-\gamma_f^2)$.  For $\tilde{y}>0$, simply take $\gamma_f\rightarrow-\gamma_f$ and we get
			\begin{equation}
				\pmat{\rho_+^{(0)}\\\rho_-^{(0)}}{1.2}=b\pmat{\delta_-\\\delta_+}{1.2}e^{-\xi_0\tilde{y}}.
			\end{equation}
			To determine the coefficients $a$ and $b$, we integrate Eq.\ (\ref{eq:0th order eq}) across the origin to obtain the condition $\delta_{\pm}^{-1}\rho_{\pm}^{(0)}(0^-)=\delta_{\mp}^{-1}\rho_{\pm}^{(0)}(0^+)$.  Note that this is just continuity of current.  We find $a=b=\mathcal{N}/2$ for some normalization $\mathcal{N}$.  
			
			The first order correction can be written as $\rho_{\pm}^{(1)}=p_{\pm}(\tilde{y})e^{i\tilde{t}}+p_{\pm}^*(\tilde{y})e^{-i\tilde{t}}=2\re\left\{p_{\pm}(\tilde{y})e^{i\tilde{t}}\right\}$, where $p_{\pm}^*$ is the complex conjugate of $p_{\pm}$.  Using this, we get
			\begin{equation}
				\frac{d}{d\tilde{y}}\left[(\pm1-\gamma_f\sgn(\tilde{y}))p_{\pm}\right]=-\gamma_{\omega}^{-1/2}(1+i\gamma_{\omega})p_{\pm}+\gamma_{\omega}^{-1/2}p_{\mp}+\frac{\gamma_{\omega}^{1/2}}{2}\frac{d\rho_{\pm}^{(0)}}{d\tilde{y}}.
				\label{eq:1st order eq}
			\end{equation}
			For $\tilde{y}<0$, 
			\begin{equation}
				\frac{d}{d\tilde{y}}\pmat{p_+\\p_-}{1.2}=\gamma_{\omega}^{-1/2}\bmat{-\delta_+(1+i\gamma_{\omega})&\delta_+\\-\delta_-&\delta_-(1+i\gamma_{\omega})}{1.2}\pmat{p_+\\p_-}{1.2}+\frac{\mathcal{N}\gamma_{\omega}^{1/2}\xi_0}{4}\pmat{\delta_+^2\\-\delta_-^2}{1.2}e^{\xi_0\tilde{y}}.
			\end{equation}
			Once again requiring $\lim\limits_{\tilde{y}\rightarrow-\infty}\rho=0$, we obtain
			\begin{equation}
				\pmat{p_+\\p_-}{1.2}=(a_+-c_+)\boldsymbol{v}_+e^{\lambda_+\tilde{y}}+(c_+\boldsymbol{v}_++c_-\boldsymbol{v}_-)e^{\xi_0\tilde{y}},
			\end{equation}
			where
			\begin{equation}
				\lambda_{\pm}=\frac{\gamma_f(1+i\gamma_{\omega})\pm\sqrt{\gamma_f^2+2i\gamma_{\omega}-\gamma_{\omega}^2}}{(1-\gamma_f^2)\gamma_{\omega}^{1/2}},
			\end{equation}
			\begin{equation}
				\boldsymbol{v}_{\pm}=\pmat{\delta_-(1+i\gamma_{\omega})-\gamma_{\omega}^{1/2}\lambda_{\pm}\\\delta_-}{1.2},
			\end{equation}
			\begin{equation}
				c_{\pm}=\mp\frac{i\mathcal{N}\gamma_f\gamma_{\omega}^{-1/2}\left(\gamma_f+i\gamma_{\omega}\pm\sqrt{\gamma_f^2+2i\gamma_{\omega}-\gamma_{\omega}^2}\right)}{4(1-\gamma_f^2)\sqrt{\gamma_f^2+2i\gamma_{\omega}-\gamma_{\omega}^2}},
			\end{equation}
			As before, take $\gamma_f\rightarrow-\gamma_f$ for $\tilde{y}>0$ and so
			\begin{equation}
				\pmat{p_+\\p_-}{1.2}=(b_--d_-)\boldsymbol{u}_-e^{\kappa_-\tilde{y}}+(d_+\boldsymbol{u}_++d_-\boldsymbol{u}_-)e^{-\xi_0\tilde{y}},
			\end{equation}
			where $b_-$ is undetermined and
			\begin{equation}
				\kappa_{\pm}=\frac{-\gamma_f(1+i\gamma_{\omega})\pm\sqrt{\gamma_f^2+2i\gamma_{\omega}-\gamma_{\omega}^2}}{(1-\gamma_f^2)\gamma_{\omega}^{1/2}},
			\end{equation}
			\begin{equation}
				\boldsymbol{u}_{\pm}=\pmat{\delta_+(1+i\gamma_{\omega})-\gamma_{\omega}^{1/2}\kappa_{\pm}\\\delta_+}{1.2},
			\end{equation}
			\begin{equation}
				d_{\pm}=\mp\frac{i\mathcal{N}\gamma_f\gamma_{\omega}^{-1/2}\left(\gamma_f-i\gamma_{\omega}\mp\sqrt{\gamma_f^2+2i\gamma_{\omega}-\gamma_{\omega}^2}\right)}{4(1-\gamma_f^2)\sqrt{\gamma_f^2+2i\gamma_{\omega}-\gamma_{\omega}^2}}.
			\end{equation}
			Integrating Eq.\ (\ref{eq:1st order eq}) gives the condition
			\begin{equation}
				\frac{1}{\delta_{\pm}}p_{\pm}(0^-)\mp\frac{\gamma_{\omega}^{1/2}}{2}\rho_{\pm}^{(0)}(0^-)=\frac{1}{\delta_{\mp}}p_{\pm}(0^+)\mp\frac{\gamma_{\omega}^{1/2}}{2}\rho_{\pm}^{(0)}(0^+).
			\end{equation}
			This system gives
			\begin{align}
				a_+=\frac{i\mathcal{N}\gamma_f\gamma_{\omega}^{1/2}\left[\gamma_f^2-\gamma_f+i\gamma_{\omega}-\gamma_{\omega}^2+(1-\gamma_f+i\gamma_{\omega})\sqrt{\gamma_f^2+2i\gamma_{\omega}-\gamma_{\omega}^2}\right]}{4(1-\gamma_f^2)\sqrt{\gamma_f^2+2i\gamma_{\omega}-\gamma_{\omega}^2}},
			\end{align}
			\begin{align}
				b_-=\frac{i\mathcal{N}\gamma_f\gamma_{\omega}^{1/2}\left[\gamma_f^2+\gamma_f+i\gamma_{\omega}-\gamma_{\omega}^2-(1+\gamma_f+i\gamma_{\omega})\sqrt{\gamma_f^2+2i\gamma_{\omega}-\gamma_{\omega}^2}\right]}{4(1-\gamma_f^2)\sqrt{\gamma_f^2+2i\gamma_{\omega}-\gamma_{\omega}^2}}.
			\end{align}
			Finally, noting that $\mathcal{N}=\rho_0(1-\gamma_f^2)$ and $\lambda_+=-\kappa_-=\xi$, we have
			\begin{align}
				\begin{split}
					p&=p_++p_-=\frac{i\rho_0\gamma_f\gamma_{\omega}^{-1/2}\sgn(\tilde{y})}{1-\gamma_f^2}\left[e^{-\xi_0\left|\tilde{y}\right|}\vphantom{\frac{1}{2}}\right.\\
					&\hspace{1in}\left.-\frac{1}{2}\left(2-\gamma_f^2+i\gamma_{\omega}+\gamma_f\sqrt{\gamma_f^2+2i\gamma_{\omega}-\gamma_{\omega}^2}\right)e^{-\xi\left|\tilde{y}\right|}\right]
				\end{split}
			\end{align}
			and
			\begin{equation}
				p_+-p_-=\frac{i\rho_0\gamma_f^2\gamma_{\omega}^{-1/2}}{1-\gamma_f^2}\left(-\frac{\xi}{\xi_0}e^{-\xi|\tilde{y}|}+e^{-\xi_0|\tilde{y}|}\right).
			\end{equation}
			
		\subsection{\label{subapp:rnt, ramp}RnT particles, ramp potential}
			Taking $\Phi'(y)=f\theta(y)$, Eq.\ (\ref{eq:rnt pde rest dim}) becomes
			\begin{equation}
				\frac{\partial\rho_{\pm}}{\partial\tilde{t}}=-\frac{\partial}{\partial\tilde{y}}\left[\left(\pm\gamma_{\omega}^{-1/2}-\epsilon\cos\tilde{t}-\gamma_f\gamma_{\omega}^{-1/2}\theta(\tilde{y})\right)\rho_{\pm}\right]-\gamma_{\omega}^{-1}\rho_{\pm}+\gamma_{\omega}^{-1}\rho_{\mp}.
			\end{equation}
			To zeroth order,
			\begin{equation}
				\frac{d}{d\tilde{y}}\left[\left(\pm1-\gamma\theta(\tilde{y})\right)\rho_{\pm}^{(0)}\right]=-\gamma_{\omega}^{-1/2}\rho_{\pm}^{(0)}+\gamma_{\omega}^{-1/2}\rho_{\mp}^{(0)}.
			\end{equation}
			For $\tilde{y}<0$,
			\begin{equation}
				\frac{d}{d\tilde{y}}\pmat{\rho_+^{(0)}\\\rho_-^{(0)}}{1.2}=\gamma_{\omega}^{-1/2}\bmat{-1 & 1\\-1 & 1}{1.2}\pmat{\rho_+^{(0)}\\\rho_-^{(0)}}{1.2}.
			\end{equation}
			The solution with $\lim\limits_{\tilde{y}\rightarrow-\infty}\rho=\rho_0$ is
			\begin{equation}
				\pmat{\rho_+^{(0)}\\\rho_-^{(0)}}{1.2}=\frac{\rho_0}{2}\pmat{1\\1}{1.2}.
			\end{equation}
			For $\tilde{y}>0$,
			\begin{equation}
				\frac{d}{d\tilde{y}}\pmat{\rho_+^{(0)}\\\rho_-^{(0)}}{1.2}=\bmat{-\delta_- & \delta_-\\-\delta_+ & \delta_+}{1.2}\pmat{\rho_+^{(0)}\\\rho_-^{(0)}}{1.2}
			\end{equation}
			where $\delta_{\pm}=1/(1\pm\gamma_f)$.  The solution satisfying $\lim_{\tilde{y}\rightarrow\infty}\rho=0$ is
			\begin{equation}
				\pmat{\rho_+^{(0)}\\\rho_-^{(0)}}{1.2}=b\pmat{\delta_-\\\delta_+}{1.2}e^{\kappa_0\tilde{y}}
			\end{equation}
			where $\kappa_0=-2\gamma_f\gamma_{\omega}^{-1/2}/(1-\gamma_f^2)$.  At $\tilde{y}=0$, the condition $\rho_{\pm}^{(0)}(0^-)=\delta_{\mp}^{-1}\rho_{\pm}(0^+)$ gives $b=\rho_0/2$.
			
			Using $\rho_{\pm}^{(1)}=p_{\pm}e^{i\tilde{t}}+p_{\pm}^*e^{i\tilde{t}}$, we have to first order
			\begin{equation}
				\frac{d}{d\tilde{y}}\Big[\left(\pm1-\gamma_f\theta(\tilde{y})\right)p_{\pm}\Big]=-\gamma_{\omega}^{-1/2}(1+i\gamma_{\omega})p_{\pm}+\gamma_{\omega}^{-1/2}p_{\mp}+\frac{\gamma_{\omega}^{1/2}}{2}\frac{d\rho^{(0)}_{\pm}}{d\tilde{y}}.
			\end{equation}
			For $\tilde{y}<0$,
			\begin{equation}
				\frac{d}{d\tilde{y}}\pmat{p_+\\p_-}{1.2}=\gamma_{\omega}^{-1/2}\bmat{-(1+i\gamma_{\omega}) & 1\\-1 & 1+i\gamma_{\omega}}{1.2}\pmat{p_+\\p_-}{1.2}.
			\end{equation}
			The solution satisfying $\lim\limits_{\tilde{y}\rightarrow-\infty}\rho=0$ is
			\begin{equation}
				\pmat{p_+\\p_-}{1.2}=\frac{\rho_0}{4}a_+\boldsymbol{v}_+e^{\lambda_+\tilde{y}},
				\label{eq:rnt, ramp, y<0}
			\end{equation}
			where $\lambda_+=\sqrt{2i-\gamma_{\omega}}$ and
			\begin{equation}
				\boldsymbol{v}_+=\pmat{1+i\gamma_{\omega}-\gamma_{\omega}^{1/2}\lambda_+\\1}{1.2}.
			\end{equation}
			For $\tilde{y}>0$, 
			\begin{equation}
				\frac{d}{d\tilde{y}}\pmat{p_+\\p_-}{1.2}=\gamma_{\omega}^{-1/2}\bmat{-\delta_-(1+i\gamma_{\omega}) & \delta_-\\-\delta_+ & \delta_+(1+i\gamma_{\omega})}{1.2}\pmat{p_+\\p_-}{1.2}+\frac{\gamma_{\omega}^{1/2}\kappa_0}{2}\pmat{\delta_-^2\\-\delta_+^2}{1.2}e^{\kappa_0\tilde{y}}.
			\end{equation}
			The solution satisfying $\lim\limits_{\tilde{y}\rightarrow\infty}\rho=0$ is
			\begin{equation}
				\pmat{p_+\\p_-}{1.2}=\frac{\rho_0}{4}(b_--d_-)\boldsymbol{u}_-e^{\kappa_-\tilde{y}}+\frac{\rho_0}{4}\Big(d_+\boldsymbol{u}_++d_-\boldsymbol{u}_-\Big)e^{\kappa_0\tilde{y}},
			\end{equation}
			where
			\begin{equation}
				\kappa_{\pm}=\frac{-\gamma_f(1+i\gamma_{\omega})\pm\sqrt{\gamma_f^2+2i\gamma_{\omega}-\gamma_{\omega}^2}}{(1-\gamma_f^2)\gamma_{\omega}^{1/2}},
			\end{equation}
			\begin{equation}
				\boldsymbol{u}_{\pm}=\pmat{\delta_+(1+i\gamma_{\omega})-\gamma_{\omega}^{1/2}\kappa_{\pm}\\\delta_+}{1.2},
			\end{equation}
			\begin{equation}
				d_{\pm}=\mp\frac{i\gamma_f\gamma_{\omega}^{-1/2}\left(\gamma_f-i\gamma_{\omega}\mp\sqrt{\gamma_f^2+2i\gamma_{\omega}-\gamma_{\omega}^2}\right)}{(1-\gamma_f^2)\sqrt{\gamma_f^2+2i\gamma_{\omega}-\gamma_{\omega}^2}}.
			\end{equation}
			At $\tilde{y}=0$, we have the condition
			\begin{equation}
				\pm p_{\pm}(0^-)-\frac{\gamma_{\omega}^{1/2}}{2}\rho_{\pm}^{(0)}(0^-)=\pm\frac{1}{\delta_{\mp}}p_{\pm}(0^+)-\frac{\gamma_{\omega}^{1/2}}{2}\rho_{\pm}^{(0)}(0^+)
			\end{equation}
			which gives
			\begin{equation}
				a_+=-\frac{\delta_-(1-\delta_-)+(1-\delta_+)\left(\delta_+(1+i\gamma_{\omega})-\gamma_{\omega}^{1/2}\kappa_-\right)-(\kappa_+-\kappa_-)d_+}{\delta_-\lambda_++\kappa_+},
			\end{equation}
			\begin{equation}
				b_-=-\frac{\delta_-(1-\delta_-)+(1-\delta_+)\delta_-\left(1+i\gamma_{\omega}-\gamma_{\omega}^{1/2}\lambda_+\right)+(\delta_-\lambda_++\kappa_-)d_+}{\delta_-\lambda_++\kappa_+}.
			\end{equation}
	
	\section{\label{app:Response Function}Expression for $B_{\omega}$}
		Using $\Delta\rho=2\epsilon\re\left\{p(y)e^{i\omega t}\right\}$, the additional force can be written as 
		\begin{equation}
			\Delta F(t)=-\int_{-\infty}^{\infty}\Phi'\Delta\rho dy=-\epsilon\left(R_{\omega}\sin\omega t+I_{\omega}\cos\omega t\right),
		\end{equation}
		where $R_{\omega}=\int2\Phi'\re\left\{ip\right\}dy$ and $I_{\omega}=\int2\Phi'\im\left\{ip\right\}dy$.  Thus, $\Delta F_{\omega'}=B_{\omega'}a_{\omega'}$ becomes
		\begin{align}
			\begin{split}
				&i\epsilon R_{\omega}\left[\delta(\omega-\omega')-\delta(\omega+\omega')\right]-\epsilon I_{\omega}\left[\delta(\omega-\omega')+\delta(\omega+\omega')\right]\\
				&\hspace{1in}=-iB_{\omega'}a\left[\delta(\omega-\omega')-\delta(\omega+\omega')\right].
			\end{split}
		\end{align}
		Matching coefficients of the delta functions, we find
		\begin{align}
			B_{\omega}&=-\frac{\epsilon}{a}(R_{\omega}+iI_{\omega}),\\
			B_{-\omega}&=-\frac{\epsilon}{a}(R_{\omega}-iI_{\omega}).
		\end{align}
		To verify that $B_{-\omega}$ is indeed the complex conjugate of $B_{\omega}$, note that the transformation $\omega\rightarrow-\omega$ takes $a\sin\omega t\rightarrow-a\sin\omega t$ or $\Delta F\rightarrow-\Delta F$, and so we must have $R_{-\omega}=R_{\omega}$ and $I_{-\omega}=-I_{\omega}$, as expected in linear response theory for in-phase and out-of-phase responses.
		\begin{equation}
			B_{\omega}=-\frac{\epsilon}{a}\left(R_{\omega}+iI_{\omega}\right)=-\frac{2i}{l_{\textrm{diffusion}}}\int_{-\infty}^{\infty}\Phi'(y)p(y)dy.
		\end{equation}
		Note that we may rewrite the force as
		\begin{equation}
			\Delta F=aB_{\omega}'\sin\omega t+aB_{\omega}''\cos\omega t.
		\end{equation}
		
	\section{\label{app:Variable Definitions}$B_{\omega}$ for RnT particles in a ramp potential}
\begin{figure}
			\centering
			\includegraphics[scale=0.4]{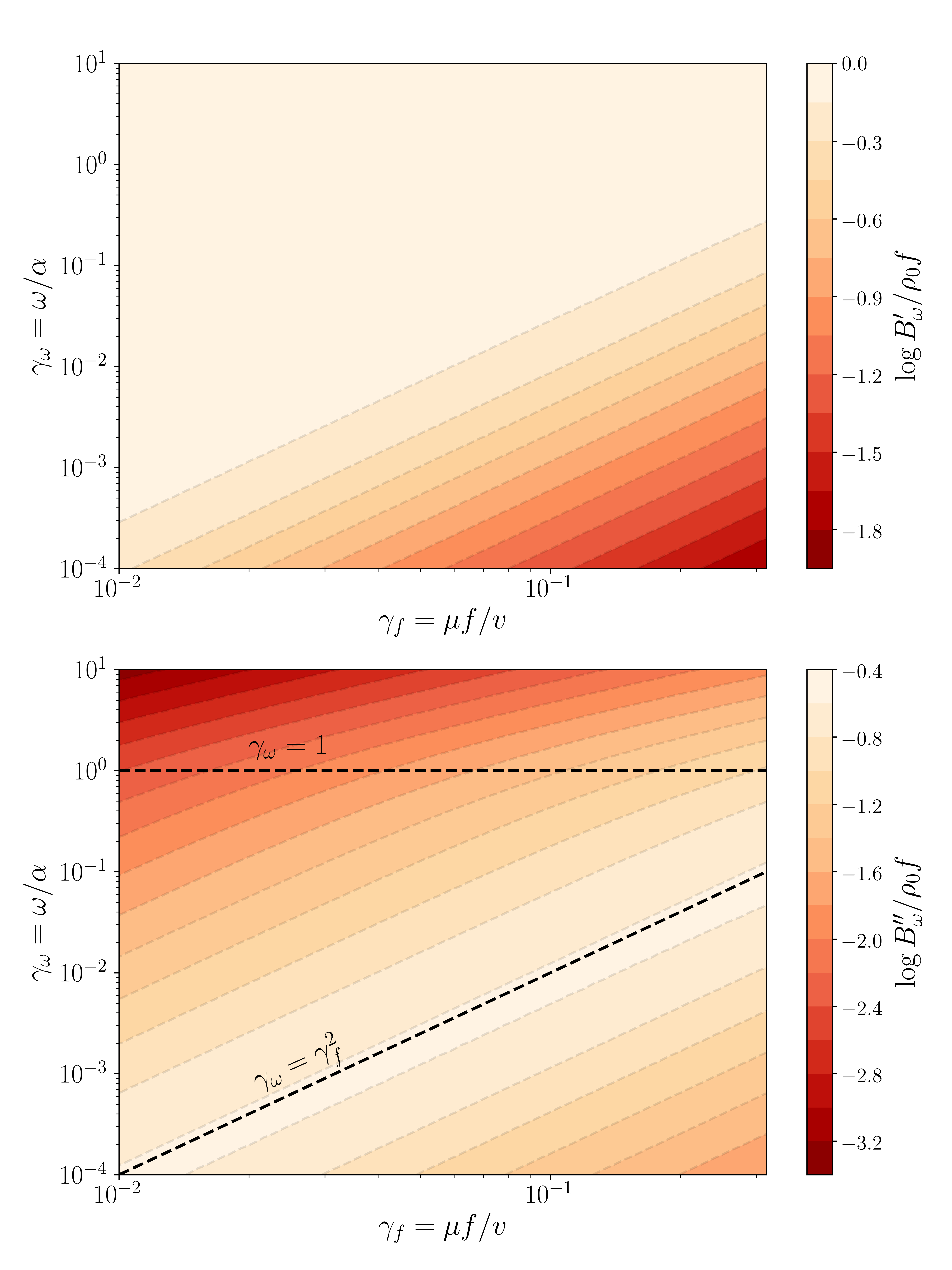}
			\caption{Level sets of $B_{\omega}$ for RnT particles in the ramp potential.  \textbf{Top:} Real part. \textbf{Bottom:} Imaginary part.  Just like in the V-shaped potential, there are two crossovers indicated by the black dashed lines: $\gamma_{\omega}\sim\gamma_f^2$, which divides the slow and passive fast regimes; and $\gamma_{\omega}\sim1$, which divides the passive fast and active fast regimes.}
			\label{fig:Bw contour ramp}
		\end{figure}

		For RnT particles in a ramp potential, the response function is
		\begin{equation}
			B_{\omega}=\frac{i\rho_0f}{2}\left[(b_--d_-)\frac{s_-}{\kappa_-}+(d_+s_++d_-s_-)\frac{1}{\kappa_0}\right],
			\label{eq:response ramp potential}
		\end{equation}
		where
		\begin{equation}
			b_-=-\frac{\delta_-(1-\delta_-)+(1-\delta_+)\delta_-\left(1+i\gamma_{\omega}-\gamma_{\omega}^{1/2}\lambda_+\right)+(\delta_-\lambda_++\kappa_-)d_+}{\delta_-\lambda_++\kappa_+}
		\end{equation}
		\begin{equation}
			d_{\pm}=\mp\frac{i\gamma_f\gamma_{\omega}^{-1/2}\left(\gamma_f-i\gamma_{\omega}\mp\sqrt{\gamma_f^2+2i\gamma_{\omega}-\gamma_{\omega}^2}\right)}{(1-\gamma_f^2)\sqrt{\gamma_f^2+2i\gamma_{\omega}-\gamma_{\omega}^2}},
		\end{equation}
		\begin{equation}
			\kappa_{\pm}=\frac{-\gamma_f(1+i\gamma_{\omega})\pm\sqrt{\gamma_f^2+2i\gamma_{\omega}-\gamma_{\omega}^2}}{(1-\gamma_f^2)\gamma_{\omega}^{1/2}},
		\end{equation}
		$s_{\pm}=\delta_+(2+i\gamma_{\omega})-\gamma_{\omega}^{1/2}\kappa_{\pm}$, $\delta_{\pm}=1/(1\pm\gamma_f)$, $\kappa_0=-2\gamma_f\gamma_{\omega}^{-1/2}/(1-\gamma_f^2)$, and $\lambda_+=\sqrt{2i-\gamma_{\omega}}$.  The level sets of $B_{\omega}$ for the ramp potential are shown in Fig.\ \ref{fig:Bw contour ramp}.
	
	\section{\label{app:Currents}Currents and dissipation}
		We here calculate the particle currents, which allows us to more easily interpret the dissipation.  The total current of passive particles relative to the static fluid (i.e.\ without the fictitious drift) is given by
		\begin{equation}
			J_x(y,t)=-\mu\Phi'(y)\rho-D\frac{\partial\rho}{\partial y}=v_x(y,t)\rho(y,t),
			\label{eq:def current}
		\end{equation}
		where $v_x(y,t)$ is the velocity field in the lab frame as a function of the co-moving coordinate $y$.  Substituting the solutions found in Appendix \ref{app:Solution Details}, we find for the V-shaped potential
		\begin{equation}
			v_x=-\epsilon\mu f\re\left\{i(\xi-\gamma)e^{-(\xi-\gamma)|\tilde{y}|}e^{i\tilde{t}}\right\},
		\end{equation}
		and for the ramp potential
		\begin{equation}
			v_x=-\epsilon\mu f\re
			\begin{cases}
				i^{3/2}\frac{(\xi-\gamma)}{\xi-\gamma+\sqrt{i}}e^{\sqrt{i}\tilde{y}}e^{i\tilde{t}}, & \tilde{y}<0\\
				i^{3/2}\frac{(\xi-\gamma)}{\xi-\gamma+\sqrt{i}}e^{-(\xi-\gamma)\tilde{y}}e^{i\tilde{t}}, & \tilde{y}>0
			\end{cases}
		\end{equation}
		where $\xi=\frac{1}{2}\left(\gamma+\sqrt{\gamma^2+4i}\right)$ and $\gamma=\mu f/\sqrt{\omega D}$ (Eq.\ (\ref{eq:def passive params})).  With this, the dissipation rate of the entire system is
		\begin{equation}
			\dot{Q}=\int_{-\infty}^{\infty}\rho(y,t)\frac{v_x(y,t)^2}{\mu}dy.
			\label{eq:Q dot integral}
		\end{equation}
		Note that if the particles have a characteristic drift velocity $V$ and a characteristic decay length for the current $L$, the average rate of dissipation goes as
		\begin{equation}
			\langle\dot{Q}\rangle\sim\rho_0L\frac{V^2}{\mu}.
		\end{equation}
		
		\subsection{\label{subapp:Currents, Slow}Slow drive}
			For slow drive, we have $\gamma\gg1$.  For the \textbf{V-shaped potential}, the velocity field is
			\begin{equation}
				v_x\approx a\omega\re\left\{e^{-\frac{1}{\gamma^3}|\tilde{y}|}e^{i\left(-\frac{1}{\gamma}|\tilde{y}|+i\tilde{t}\right)}\right\}.
			\end{equation}
			The characteristic velocity is $V\sim a\omega$.  Note that since $\rho=\rho_0e^{-\gamma|\tilde{y}|}$, variations in the velocity field occur on length scales much greater than the decay length of the density $L\sim D/\mu f$.  To leading order, the dissipation rate is
			\begin{equation}
				\dot{Q}_{\textrm{V-shaped}}=\int_{-\infty}^{\infty}\rho\frac{v_x^2}{\mu}dy\approx\rho_0\frac{2D}{\mu f}\frac{(a\omega)^2}{\mu}\cos^2\omega t.
			\end{equation}
			Time averaging, we have
			\begin{equation}
				\langle\dot{Q}\rangle_{\textrm{V-shaped}}\approx\rho_0\frac{D}{\mu f}\frac{(a\omega)^2}{\mu}.
			\end{equation}
			For the \textbf{ramp potential}, the velocity field is
			\begin{equation}
				v_x\approx a\omega\re
				\begin{cases}
					e^{\sqrt{i}\tilde{y}}e^{i\tilde{t}}, & \tilde{y}<0\\
					e^{-\frac{1}{\gamma^3}\tilde{y}}e^{i\left(-\frac{1}{\gamma}\tilde{y}+i\tilde{t}\right)}, & \tilde{y}>0
				\end{cases}
			\end{equation}
			The characteristic velocity is $V\sim a\omega$.  In this case, the dominant contribution to the dissipation will be from the bulk $\tilde{y}<0$ since $l_{\textrm{diffusion}}\gg l_{\textrm{penetration}}$.  The dissipation rate is
			\begin{equation}
				\dot{Q}_{\textrm{ramp}}\approx\frac{1}{4\sqrt{2}}\rho_0\sqrt{\frac{D}{\omega}}\frac{(a\omega)^2}{\mu}(2+\cos2\omega t+\sin2\omega t).
			\end{equation}
			The average dissipation is then
			\begin{equation}
				\langle\dot{Q}\rangle_{\textrm{ramp}}\approx\frac{1}{2\sqrt{2}}\rho_0\sqrt{\frac{D}{\omega}}\frac{(a\omega)^2}{\mu}.
			\end{equation}
			The average dissipation rates only differ by multiplication by a constant.
			
		\subsection{\label{subapp:Currents, Fast}Fast drive, diffusive particles}
			For fast drive, $\gamma\ll1$.  The velocity field for the \textbf{V-shaped potential} is
			\begin{equation}
				v_x\approx a\omega\gamma e^{-|\tilde{y}|}\sin\left(-|\tilde{y}|+\omega t+\frac{\pi}{4}\right).
			\end{equation}
			For the \textbf{ramp potential}, the velocity field is
			\begin{equation}
				v_x\approx\frac{1}{2}a\omega\gamma e^{-|\tilde{y}|}\sin\left(-|\tilde{y}|+\omega t+\frac{\pi}{4}\right).
			\end{equation}
			The factor of $1/2$ is due to the diffusive flux (Sec.\ \ref{subsec:Fast drive}).  In this fast drive case, the decay length of the velocity field for both potentials dominates over the decay length of the density; we can treat the density as constant over this distance.  Thus,
			\begin{align}
				\dot{Q}\approx\rho_0\int_{-\infty}^{\infty}\frac{v_x^2}{\mu}dy\approx\left(\frac{1}{4}\right)\frac{1}{\sqrt{2}}\rho_0\sqrt{\frac{D}{\omega}}\frac{(a\omega\gamma)^2}{\mu}\left[1+\frac{1}{\sqrt{2}}\sin\left(2\omega t-\frac{\pi}{4}\right)\right],
			\end{align}
			or
			\begin{equation}
				\langle\dot{Q}\rangle\approx\left(\frac{1}{4}\right)\frac{1}{\sqrt{2}}\rho_0\sqrt{\frac{D}{\omega}}\frac{(a\omega\gamma)^2}{\mu},
			\end{equation}
			where the factor of $1/4$ corresponds to the ramp potential.
			
		\subsection{\label{subapp:Currents, Persistent}Fast drive, persistent particles}
			For RnT particles, the current is instead given by
			\begin{equation}
				J_x(y,t)=J_++J_-=(\langle v\rangle-\mu f\sgn(y))\rho,
			\end{equation}
			where $\langle v\rangle=v(\rho_+-\rho_-)/\rho$ and $\rho=\rho_++\rho_-$.  The particles responsible for the current have a velocity field
			\begin{equation}
				v_x(y,t)=\frac{J_x}{\rho}=\langle v\rangle-\mu f\sgn(y).
			\end{equation}
			Substituting the solution for the V-shaped potential, we obtain
			\begin{align}
				v_x=\epsilon\mu f\frac{\gamma_f}{\gamma_{\omega}^{1/2}}\re\left\{i\left(1-\sqrt{1+2i\frac{\gamma_{\omega}}{\gamma_f^2}-\frac{\gamma_{\omega}^2}{\gamma_f^2}}\right)e^{-(\xi-\xi_0)|\tilde{y}|}e^{i\omega t}\right\}.
			\end{align}
			Taking $\gamma_{\omega}\gg1$ or $\omega\gg\alpha$, we get
			\begin{equation}
				v_x\approx a\omega\gamma_f e^{-\gamma_{\omega}^{-1/2}|\tilde{y}|}\cos(-\gamma_{\omega}^{1/2}|\tilde{y}|+\omega t),
			\end{equation}
			which using Eq.\ (\ref{eq:Q dot integral}) gives
			\begin{equation}
				\langle\dot{Q}\rangle_{\textrm{V-shaped}}\approx\rho_0\frac{a^2\mu f^2\alpha}{v}\left(\frac{\omega}{\alpha}\right)^2.
			\end{equation}
			The expression of velocity for the ramp potential is cumbersome.  However, taking $\gamma_{\omega}\gg1$, we find
			\begin{equation}
				\langle\dot{Q}\rangle_{\textrm{ramp}}\approx\frac{1}{4}\rho_0\frac{a^2\mu f^2\alpha}{v}\left(\frac{\omega}{\alpha}\right)^2.
			\end{equation}
			The dissipation rates only differ by multiplication by a constant.


\begin{thebibliography}{10}
		\bibitem{Marchetti et al}
			M. C. Marchetti, J. F. Joanny, S. Ramaswamy, T. B. Liverpool, J. Prost, M. Rao, R. A. Simha, Hydrodynamics of soft active matter, Rev. Mod. Phys. \textbf{85} 1143 (2013)
		\bibitem{Bechinger et al}
			C. Bechinger, R. Di Leonardo, H. L\"{o}wen, C. Reichhardt, G. Volpe, G. Volpe, Active particles in complex and crowded environments, Rev. Mod. Phys. \textbf{88} 045006 (2016)
		\bibitem{Berg et Brown}
			H. C. Berg, D. A. Brown, Chemotaxis in Escherichia coli analysed by three-dimensional tracking, Nature \textbf{239} 500 (1972)
		\bibitem{Palacci et al}
			J. Palacci, S. Sacanna, S. H. Kim, G. R. Yi, D. J. Pine, P. M. Chaikin, Light-activated self-propelled colloids, Phil. Trans. Royal Soc. A: Math. Phys. \textbf{372} 20130372 (2014)
		\bibitem{Paxton et al}
			W. F. Paxton, K. C. Kistler, C. C. Olmeda, A. Sen, S. K. St. Angelo, T. Cao, T. E. Mallouk, P. E. Lammert, V. H. Crespi, Catalytic nanomotors: Autonomous movement of striped nanorods, J. Am. Chem. Soc. \textbf{126} 13424 (2004)
		\bibitem{Astumian}
			R. D. Astumian, Thermodynamics and kinetic of a Brownian motor, Science \textbf{276} 917 (1997)
		\bibitem{Julicher et al}
			F. J\"{u}licher, A. Ajdari, J. Prost, Modeling molecular motors, Rev. Mod. Phys. \textbf{69} 1269 (1997)
		\bibitem{Cates and Tailleur}
			M. E. Cates, J. Tailleur, Motility-induced phase separation, Annu. Rev. Condens. Matter Phys. \textbf{6} 219 (2015)
		\bibitem{Redner et al}
			G. S. Redner, M. F. Hagan, A. Baskaran, Structure and dynamics of a phase-separating active colloidal fluid, Phys. Rev. Lett. \textbf{110} 055701 (2013)
		\bibitem{Fily and Marchetti}
			Y. Fily, M. C. Marchetti, Athermal phase separation of self-propelled particles with no alignment, Phys. Rev. Lett. \textbf{108} 235702 (2012)
		\bibitem{Wysocki et al}
			A. Wysocki, R. G. Winkler, G. Gompper, Cooperative motion of active Brownian spheres in three-dimensional dense suspensions, EPL \textbf{105} 48004 (2014)
		\bibitem{Bialke et al}
			J. Bialk\'{e}, H. L\"{o}wen, T. Speck, Microscopic theory for the phase separation of self-propelled repulsive disks, EPL \textbf{103} 30008 (2013)
		\bibitem{Galajda et al}
			P. Galajda, J. Keymer, P. M. Chaikin, R. Austin, A wall of funnels concentrates swimming bacteria, J. Bacteriol. \textbf{189} 8704 (2007)
		\bibitem{Di Leonardo et al}
			R. Di Leonardo, L. Angelani, D. Dell'Arciprete, G. Ruocco, V. Lebba, S. Schippa, M. P. Conte, F. Mecarini, F. De Angelis, E. Di Fabrizio, Bacterial ratchet motors, Proc. Natl. Acad. Sci. USA \textbf{107} 9541 (2010)
		\bibitem{Vizsnyiczai et al}
			G. Vizsnyiczai, G. Frangipane, C. Maggi, F. Saglimbeni, S. Bianchi, R. Di Leonardo, Light controlled 3D micromotors powered by bacteria, Nat. Comm. \textbf{8} 15874 (2017)
		\bibitem{Sokolov et al}
			A. Sokolov, I. S. Aranson, Reduction of viscosity in suspension of swimming bacteria, Phys. Rev. Lett. \textbf{103} 148101 (2009)
		\bibitem{Rafai et al}
			S. Rafa\"{i}, L. Jibuti, P. Peyla, Effective viscosity of microswimmer suspensions, Phys. Rev. Lett. \textbf{104} 098102 (2010)
		\bibitem{Hatwalne et al}
			Y. Hatwalne, S. Ramaswamy, M. Rao, R. A. Simha, Rheology of active-particles suspensions, Phys. Rev. Lett. \textbf{92} 118101 (2004)
		\bibitem{Turlier et al}
			H. Turlier, D. A. Fedosov, B. Audoly, T. Auth, N. S. Gov, C. Sykes, J. F. Joanny, G. Gompper, T. Betz, Equilibrium physics breakdown reveals the active nature of red blood cell flickering, Nat. Phys. \textbf{12} 513 (2016)
		\bibitem{Chu et al}
			F. Y. Chu, S. C. Haley, A. Zidovska, On the origin of shape fluctuations of the cell nucleus, Proc. Natl. Acad. Sci. USA \textbf{114} 10338 (2017)
		\bibitem{Fodor et al}
			\'{E}. Fodor, M. Guo, N. S. Gov, P. Visco, D. A. Weitz, F. van Wijland, Activity-driven fluctuations in living cells, EPL \textbf{110} 48005 (2015)
		\bibitem{Bi et al}
			D. Bi, X. Yang, M. C. Marchetti, M. L. Manning, Motility-driven glass and jamming transitions in biological tissues, Phys. Rev. X \textbf{6} 021011 (2016)
		\bibitem{Caprini et al}
			L. Caprini, U. M. B. Marconi, A. Vulpiani, Linear response and correlation of a self-propelled particle in the presence of external fields, J. Stat. Mech. \textbf{2018} 033203 (2018)
		\bibitem{Solon et al}
			A. P. Solon, Y. Fily, A. Baskaran, M. E. Cates, Y. Kafri, M. Kardar, J. Tailleur, Pressure is not a state function for generic active fluids, Nat. Phys. \textbf{11} 673 (2015)
		\bibitem{Marconi et al}
			U. M. B. Marconi, A. Sarracino, C. Maggi, A. Puglisi, Self-propulsion against a moving membrane: Enhanced accumulation and drag force, Phys. Rev. E \textbf{96} 032601 (2017)
		\bibitem{Sheshka et al}
			R. Sheshka, P. Recho, L. Truskinovsky, Rigidity generation by nonthermal fluctuations, Phys. Rev. E \textbf{93} 052604 (2016)
		\bibitem{Schnitzer}
			M. J. Schnitzer, Theory of continuum random walks and application to chemotaxis, Phys. Rev. E \textbf{48} 2553 (1993)
		\bibitem{Fourier French}
			Joseph Fourier, Th\'{e}ory Analytique de la Chaleur, Chez Firmin Didot, Paris (1822)
		\bibitem{Fourier English}
			Joseph Fourier, Analytical Theory of Heat, Cambridge University Press, Cambridge (1878)
	\end{thebibliography}
\end{document}